\documentclass[a4paper,11pt]{article}
\usepackage{pos}
\usepackage{enumitem}

\newtheorem{thm}{Theorem}[section]
\newtheorem{prop}[thm]{Proposition}
\newtheorem{defn}[thm]{Definition}

\newtheorem{example}[thm]{Example}
\newenvironment{eg}{\begin{example}  \rm  }{    \end{example} }

\newtheorem{exercise}[thm]{Exercise}

\newcommand{\eop}{{\hspace*{\fill}$\square$}}

\renewcommand{\Ref}[1]{Ref.~\cite{#1}}
\newcommand{\Fig}[1]{Fig.~\ref{#1}}
\newcommand{\Eq}[1]{Eq.~\eqref{#1}}
\newcommand{\Eqs}[2]{Eqs.~\eqref{#1} and \eqref{#2}}
\newcommand{\Sec}[1]{Sec.~\ref{#1}}

\DeclareMathOperator{\Tr}{Tr}
\DeclareMathOperator{\Span}{span}

\newcommand{\bra}[1]{\langle #1|}
\newcommand{\ket}[1]{|#1\rangle}
\newcommand{\braket}[2]{\langle #1|#2\rangle}
\newcommand{\ketbra}[2]{| #1 \rangle \langle #2 |}
\newcommand{\rent}[2]{D( #1 \, \Vert \, #2 )}

\newcommand{\Hil}{{\cal H}}
\newcommand{\Oh}{{\mathcal{O}}}
\newcommand{\Lin}{{\mathcal{L}}}

\newcommand{\mrm}[1]{\mathrm{#1}}

\newcommand{\Note}{\noindent {\bf Note:~}}

\title{Modave Lectures on Quantum Information\\{\normalsize An Introduction to Channels and Applications to Black Holes and AdS/CFT}}
\ShortTitle{Modave Lectures on Quantum Information}

\author*{Aidan Chatwin-Davies}

\affiliation{KU Leuven, Institute for Theoretical Physics\\
 Celestijnenlaan 200D, B-3001 Leuven, Belgium}

\emailAdd{aechatwi@gmail.com}

\abstract{These notes introduce a handful of core ideas from quantum information science that figure prominently in modern research on quantum gravity.
The central concept that forms the base of these notes is that of a quantum channel; that is, the most general physically-reasonable map between quantum states and between operators on Hilbert space.
After reviewing some fundamentals, we will study channels and their properties, and then go on to formulate quantum error correction in terms of quantum channels.
Along the way, we will see how a handful of problems in high energy physics, such as the black hole information problem and bulk reconstruction in AdS/CFT, can be cast in the information-theoretic language being set up.}

\FullConference{%
  XVI Modave Summer School in Mathematical Physics - Modave 2020\\
  9-11 September 2020\\
  Brussels, Belgium
}

\tableofcontents

\begin{document}
\maketitle

\section{Introduction}

Quantum Information Science (QIS) sits at an intersection point of physics, mathematics, and computer science.
The field concerns itself with the information contained in quantum mechanical systems, how that information can be encoded, manipulated, and retrieved, and how these operations' properties, capabilities, and limitations can be quantified.
As we sit at the cusp of the era of quantum computers, the practical importance of QIS only continues to increase.
In parallel, QIS continues to drive new discoveries and further our theoretical understanding of questions in high energy physics.

The aim of these notes is to explain a handful of core ideas from QIS that figure prominently in modern research on quantum gravity.
They are certainly not a complete introduction to QIS nor its application to gravity; nevertheless, they will hopefully be both interesting and useful for someone who wants to learn a bit more about the information theory that underpins gravitational applications.
These notes should be accessible to anyone with a solid command of undergraduate quantum physics.

Many parts of these notes are based on my own experiences learning about QIS as a student, and as such are heavily inspired by John Preskill's excellent set of lecture notes \cite{preskill}.
Other parts draw on Mark Wilde's comprehensive text on quantum Shannon theory \cite{Wilde:2011npi}.
In these parts and elsewhere, I will point the reader to original source material when available, as well as to further reading.

So, what is quantum information?
The abstract and somewhat tautological answer is that it is the information contained in the state of a quantum mechanical system.
It's not very illuminating, not to mention that we could give an analogously impractical definition for classical information.
However, much as we can characterize classical information science concretely as the study and manipulation of bit strings,
\begin{equation}
x_1 x_2 \cdots x_n \qquad x_i \in \{0,1\} ~~ \text{for} ~~ 1 \leq i \leq n,
\end{equation}
we can similarly characterize quantum information science as the study and manipulation of \emph{qubit} strings,
\begin{equation} \label{eq:qubit_string}
\sum_{x_1 \in \{0,1\}} \, \sum_{x_2 \in \{0,1\}} \cdots \sum_{x_n \in \{0,1\}} c_{x_1 x_2 \cdots x_n} \ket{x_1} \otimes \ket{x_2} \otimes \cdots \otimes \ket{x_n},
\end{equation}
where each orthonormal set $\{\ket{x_i = 0}, \ket{x_i = 1}\}$ spans a two-dimensional Hilbert space, $c_{x_1 x_2 \cdots x_n} \in \mathbb{C}$ for $1 \leq i \leq n$, and
\begin{equation}
\sum_{x_1 \in \{0,1\}} \, \sum_{x_2 \in \{0,1\}} \cdots \sum_{x_n \in \{0,1\}} |c_{x_1 x_2 \cdots x_n}|^2 = 1.
\end{equation}
If we can think of classical information at a concrete level as bit strings, then a concrete way to think of quantum information is as qubit strings.

A perhaps more illuminating question to ask is how quantum information and the quantum systems that store it differ from their classical counterparts.
For starters:
\begin{itemize}
\item \emph{Quantum systems exhibit true randomness.}

We can of course simulate randomness with a classical computer and use it as a resource for computation, yet such processes are fundamentally only pseudo-random.
In contrast, the outcomes of indefinite quantum measurements are truly random, at least according to the conventional pragmatic viewpoint \cite{Bell}.

\item \emph{Quantum information cannot be cloned.}

There are no fundamental barriers to making copies of a given bit string, even if the string is unknown---a photocopier copies regardless of the input.
However, the no-cloning theorem says otherwise for quantum states.
There exists no unitary process that lets one make a copy of an arbitrary, unknown state.
(See, e.g. \cite[Chap.~12.3]{griffiths_schroeter_2018}.)

\item \emph{Uncertainty limits information retrieval.}

Many quantum observables fail to commute.
This places limits on the information that can be simultaneously retrieved from a state.

\item \emph{Components of a quantum system can be entangled.}

Quantum systems can store information nonlocally.
An analogy is as follows: If classical, local information is the content of the pages in a book, nonlocal information would be information stored in correlations among the pages.
In particular, you need all of the pages in order to access the nonlocal information.
These correlations are so strong that the quantum book's pages are altered after having been read, so reading a single page at a time generally ruins the nonlocal information.

\item \emph{Quantum states can exist in superpositions.}

A common platitude is that the ability to manipulate qubit strings is so powerful because they have exponentially many states.
While this counting is correct---the dimension of the Hilbert space of $n$ qubits is $2^n$---it is also true that one can form $2^n$ different strings out of $n$ bits.
Rather, what makes operations on qubit strings special is that their states can be superpositions, as in \Eq{eq:qubit_string}.

\end{itemize}

It turns out that these differences can be exploited to perform tasks that are surprising from a classical standpoint.
For example, given a large positive integer that is the product of two large prime numbers, superposition may be used in a clever way to find the prime factors exponentially faster than the best known methods using a classical computer that processes bit strings.
This is Shor's factoring algorithm \cite{Shor:1994jg}.
Another example is the process known as quantum teleportation \cite{PhysRevLett.70.1895}, in which entanglement shared between (possibly distant) parties can be used to faithfully transfer an arbitrary quantum state from one party to the other without explicitly transporting any physical qubits.

A device that manipulates qubits to perform computations is called a \emph{quantum computer}.
The design of interesting algorithms that can run on quantum computers, as well as the task of actually building such devices are some of the more practical aspects of QIS.
While we will not spend much time on these topics, an introduction to QIS would be somewhat askew without mention of them, so let's at least sketch what a quantum computation is at a schematic level.

\begin{figure}[ht]
\centering
\includegraphics[scale=1]{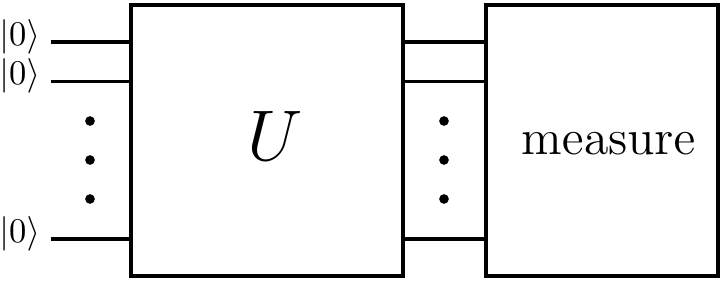}
\caption{A quantum computation, schematically.}
\label{fig:circuit}
\end{figure}

A quantum computation essentially consists of three steps, as depicted in \Fig{fig:circuit}.
First, a quantum computer that implements a state space consisting of some number of qubits, $n$, is initialized to a known initial state, say $\ket{0}^{\otimes n}$.
Next, the ``computation'' itself consists of some unitary operation, $U$, that gets applied to the $n$ qubits.\footnote{Invariably, $U$ is built out of a sequence of simpler unitary operations, or \emph{gates}, that act on smaller numbers of qubits. A collection of gates that can approximate any unitary acting on $n$ qubits arbitrarily well is called a \emph{universal gate set}. See \cite[Chap.~6]{preskill} for more details.}
In the last step, the final state is measured in the computational basis, i.e., the qubit basis $\{ \ket{x_1}\otimes \ket{x_2} \otimes \cdots \otimes \ket{x_n} ~ | ~ x_i \in \{0,1\}~\text{for}~1 \leq i \leq n \}$.
The result is that we end up sampling the probability distribution
\begin{equation}
\text{Pr}(x_1,x_2, \dots, x_n) = |\bra{x_1} \otimes \bra{x_2} \otimes \cdots \otimes \bra{x_n} ~ U ~ \ket{0} \otimes \ket{0} \otimes \cdots \otimes \ket{0}|^2.
\end{equation}
Hopefully, a bit string that encodes the answer to an interesting problem occurs with high probability!
Shor's algorithm is an example of a quantum computation.
Deutsch's algorithm is a simpler introductory example, which you can find explained anywhere from Preskill's notes \cite[Chap.~1]{preskill} to Wikipedia.

In addition to quantum algorithms and physical device implementations, a crucial ingredient for quantum computation is \emph{quantum error correction}.
In implementing a given unitary $U$ on a quantum computer, we are bound to make small errors along the way.
Moreover, even if we never made any errors in implementation, we can never perfectly isolate the qubits inside the computer from the rest of the universe.
Unwanted interactions with external degrees of freedom (like the physical components of the computer, cosmic microwave background photons, etcetera) cause the computer's qubits to bleed information into the external environment, leading to decoherence of its computational state.
It's clear that we need schemes to protect computations from these types of noise and to correct errors when they occur.

Unlike algorithms and implementations, quantum error correction \emph{is} a topic that we will take up in these notes.
We will look at an example of a quantum error correcting code as a means of introducing the subject, but we will also investigate general information-theoretic features of quantum error correction.
It turns out that this will lead to interesting applications in holography.

\emph{A posteriori}, such a connection may not be so surprising because quantum information is universal.
In a sense, all quantum systems process quantum information.
While this observation naturally leads to practical applications in the case of quantum computers, applying information-theoretic tools and techniques to other quantum phenomena can result in some considerable theoretical mileage.

The core idea that will form the base of our studies here is that of a \emph{quantum channel}.
A quantum channel is the most general, physically-reasonable map between quantum states.
Quantum channels therefore describe the most general way that a quantum system can evolve, and so, when applied to specific systems and circumstances, channels' information-theoretic properties are a powerful tool for understanding how systems process quantum information.

In \Sec{sec:QIbasics}, we will begin by reviewing some basic concepts in quantum information science, including the indispensable quantity called Von Neumann entropy.
Next, we will carefully define quantum channels in \Sec{sec:channels} and examine some of their most important properties.
\Sec{sec:QEC} is devoted to quantum error correction.
In the first part, we will see an example of a simple quantum error-correcting code, and in the second part, we will cast quantum error correction in the language of quantum channels.
Finally, in \Sec{sec:holo}, we will see how all of the tools that we will have built up can be applied to the Anti de Sitter/Conformal Field Theory correspondence to understand how localized quantum gravitational degrees of freedom are encoded in the dual quantum field-theoretic description.

\section{Quantum information basics}
\label{sec:QIbasics}

This section reviews some elementary concepts in quantum mechanics, such as states and tensor products, as well as some elementary concepts in quantum information science, such as Von Neumann entropy and relative entropy.
An experienced reader could easily skip over this section, although it may be useful to refer back to for checking conventions.

\subsection{States and multipartite Hilbert spaces}

Let us begin by defining pure and mixed states to establish some notation.

\begin{defn}
Let $\Hil$ be a Hilbert space with dimension $\dim \Hil = d$, and let $\{ \ket{i} \}_{i=1}^d$ be an orthonormal basis for $\Hil$.
Denote the space of linear operators on $\Hil$ by $\mathcal{L}(\Hil)$.
\begin{itemize}
\item A \emph{pure state} $\ket{\psi} \in \Hil$ is a normalized element of $\Hil$, to wit,
\begin{equation*}
\ket{\psi} = \sum_{i=1}^d c_i \ket{i} \quad \text{\emph{for some}} ~ c_i \in \mathbb{C}, \quad \text{\emph{and}} \quad \braket{\psi}{\psi} = \sum_{i=1}^d |c_i|^2 = 1.
\end{equation*}
\item A \emph{mixed state} $\rho \in \mathcal{L}(\Hil)$, also called a \emph{density operator} or \emph{density matrix}, is a Hermitian, positive semi-definite linear operator with unit trace, to wit,
\begin{equation*}
\rho = \sum_{i,j=1}^d \rho_{ij} \ketbra{i}{j} \quad \text{\emph{for some}} ~ \rho_{ij} \in \mathbb{C}, \quad \rho_{ij} = \rho_{ji}^*, \quad \sum_{i=1}^d \rho_{ii} = 1, \quad \bra{\psi}\rho\ket{\psi} \geq 0 ~~\forall~\ket{\psi}\in \Hil.
\end{equation*}
\end{itemize}
\end{defn}
\Note While it's fine if $d$ is countably infinite in the definition above, in the rest of these notes we will always work with finite-dimensional Hilbert spaces unless explicitly indicated.

~

\Note We will denote the set of density operators on a Hilbert space $\Hil$ by ${\cal S}(\Hil)$. 

~

\noindent For convenience, let's collect some essential properties of density operators:
\begin{enumerate}
\item $\rho = \rho^\dagger$ (density operators are Hermitian)
\item $\bra{\psi} \rho \ket{\psi} \geq 0$ for all $\ket{\psi} \in \Hil$ (density operators are positive semi-definite)
\item $\Tr \rho = 1$ (normalization)
\item Given $\rho$, there exists an orthonormal basis $\{ \ket{p_a} \}_{a=1}^d$ such that
\begin{equation*}
\rho = \sum_{a} p_a \ketbra{p_a}{p_a}, \quad p_a \geq 0, \quad \sum_a p_a = 1.
\end{equation*}
\item $\rho$ is pure if and only if one $p_a$ is nonzero and equal to 1, in which case $\rho = \ketbra{p_a}{p_a}$.
\item If $\{\Lambda_k\}_{k = 1}^{K}$ is a complete set of projectors (where $\sum_{i=1}^K \Lambda_i = I$) describing a set of measurement outcomes, the probability of obtaining outcome $i$ is $\Tr(\rho \Lambda_i)$.
\item The expectation value of an operator $\Oh \in \Lin(\Hil)$ is given by $\langle \Oh \rangle = \Tr(\rho \Oh)$.
\end{enumerate}

\noindent Next, recall the joint description of a Hilbert space with several factors:

\begin{defn}
Given two Hilbert spaces $\Hil_A$ and $\Hil_B$ with orthonormal bases $\{ \ket{i}_A \}_{i=1}^{d_A}$ and $\{ \ket{\mu}_B \}_{\mu=1}^{d_B}$, respectively, the \emph{joint Hilbert space} is denoted by $\Hil_{AB} \equiv \Hil_A \otimes \Hil_B$. $\Hil_{AB}$ has dimension $d_{AB} = d_A d_B$, and an orthonormal basis is $\{\ket{i}_A \otimes \ket{\mu}_B \}_{i=1, \mu = 1}^{d_A, d_B}$.
\end{defn}

\noindent In particular, we can always expand a state $\ket{\psi}_{AB} \in \Hil_{AB}$ as
\begin{equation}
\ket{\psi}_{AB} = \sum_{i = 1}^{d_A} \sum_{\mu = 1}^{d_B} c_{i\mu} \ket{i}_A \otimes \ket{\mu}_B.
\end{equation}
We will often omit the tensor product symbol for brevity, and we will sometimes concatenate multiple kets together when the meaning is clear.
Specifically, $\ket{i}_A \otimes \ket{\mu}_B$, $\ket{i}_A \ket{\mu}_B$, and $\ket{i\mu}_{AB}$ are all equivalent.

The last elementary ingredient that we need to recall is the partial trace.
While the tensor product lets us build a composite Hilbert space out of two factors, the partial trace lets us reduce an operator defined on a composite Hilbert space to an operator acting on a single factor.
Given $\Hil_{AB}$, suppose that we want to reduce to $\Hil_A$.
We can construct the partial trace by viewing the bra $\bra{\mu}_B$, which originally denotes the linear functional on $\Hil_B$ dual to $\ket{\mu}_B$, as an isometry $\bra{\mu}_B : \Hil_{AB} \rightarrow \Hil_{A}$ whose action is defined in terms of an orthonormal basis as
\begin{equation}
\bra{\mu}_B \left(\ket{i}_A \otimes \ket{\nu}_B\right) = \ket{i}_A \braket{\mu}{\nu}_B = \delta_{\mu \nu} \ket{i}_A.
\end{equation}

\begin{defn}
The \emph{partial trace} with respect to $B$ is the linear map $\Tr_B : \Lin(\Hil_{AB}) \rightarrow \Lin(\Hil_A)$ whose action on an operator $\Oh_{AB}$ is given in terms of an orthonormal basis of $\Hil_B$, $\{\ket{\mu}_B\}_{\mu = 1}^{d_B}$, by
\begin{equation*}
\Tr_B \Oh_{AB} = \sum_{\mu = 1}^{d_B} {}_B\bra{\mu} \Oh_{AB} \ket{\mu}_B .
\end{equation*}
The action of the resulting operator $\Oh_A \equiv \Tr_B \Oh_{AB}$ on a state $\ket{\psi}_A \in \Hil_A$ is given by
\begin{equation*}
\Oh_A \ket{\psi}_A = \sum_{\mu = 1}^{d_B} {}_B\bra{\mu} \left( \Oh_{AB} (\ket{\psi}_A \otimes \ket{\mu}_B) \right).
\end{equation*}
\end{defn}
\Note Of course, we can easily interchange $A$ and $B$ so that we reduce to the factor $B$ (or ``trace out'' $A$) instead.

~

The partial trace is a way to implement ignorance about a factor of a multipartite Hilbert space.
For example, if we only have access to a single part $A$ of a larger Hilbert space, then a partial trace over the complement of $A$ reveals how states appear and how operators act when restricted to $\Hil_A$ alone.

\begin{eg}
Let $\rho_{AB}$ be a density operator on $\Hil_{AB}$, which we write in terms of orthonormal bases for $\Hil_A$ and $\Hil_B$ as
\begin{equation}
\rho_{AB} = \sum_{i, \mu} \sum_{j, \nu} \rho_{i\mu \, j\nu} \ket{i}_A\ket{\mu}_B \bra{j}_A \bra{\nu}_B .
\end{equation}
Taking the partial trace with respect to $B$ gives us the reduced state on $A$:
\begin{align*}
\rho_A \equiv \Tr_B \rho_{AB} &= \sum_{\lambda} \bra{\lambda}_B \left( \sum_{i, \mu} \sum_{j, \nu} \rho_{i\mu \, j\nu} \ket{i}_A\ket{\mu}_B \bra{j}_A \bra{\nu}_B \right) \ket{\lambda}_B \\
&= \sum_{i , \mu} \sum_{j , \nu} \rho_{i\mu \, j\nu} \ketbra{i}{j}_A \left( \sum_\lambda \braket{\lambda}{\mu}_B \braket{\nu}{\lambda}_B \right) \\
&= \sum_{i , \mu} \sum_{j , \nu} \rho_{i\mu \, j\nu} \ketbra{i}{j}_A ~ \delta_{\mu\nu} \\
&= \sum_{i,j} \left(\sum_{\mu} \rho_{i\mu \, j \mu} \right) \ketbra{i}{j}_A
\end{align*}
As an exercise, you can check that $\rho_A$ is Hermitian, positive semi-definite, and normalized.

\eop
\end{eg}

\subsection{Von Neumann entropy}

Von Neumann Entropy is a quantity of singular importance for quantum information.
Its definition is as follows.

\begin{defn}
The \emph{Von Neumann entropy} of a state $\rho \in \mathcal{S}(\Hil)$, denoted by $S(\rho)$, is
\begin{equation}
S(\rho) = - \Tr \left( \rho \log \rho \right).
\end{equation}
\end{defn}
\Note This definition also holds for infinite-dimensional Hilbert spaces.

~

For example, if we write a state $\rho$ in its eigenbasis as $\rho = \sum_i p_i \ketbra{p_i}{p_i}$, then its Von Neumann entropy is\footnote{If you are familiar with classical information theory, then you might notice that this coincides with the classical Shannon entropy of the probability distribution $\{p_i\}$.
We will not go into classical information theory in these notes beyond this remark, but I encourage you to take a look at Claude Shannon's original manuscripts, which are a concise and accessible introduction to the subject \cite{shannon}.
Likewise, Wilde's text \cite{Wilde:2011npi} gives a thorough and positioned account of classical information theory as a precursor to quantum information theory.}
\begin{equation}
S(\rho) = - \sum_i p_i \log p_i .
\end{equation}
We also tacitly take $p_i \log p_i$ to be continuous at $p_i = 0$, taking the value 0.
In particular, this means that $S(\rho) = 0$ if $\rho = \ketbra{\psi}{\psi}$ is a pure state.

We now list several properties of Von Neumann entropy that will help us to interpret it.
Again, we assume that $\dim \Hil = d < \infty$.

\begin{prop} Some properties of Von Neumann entropy:
\begin{enumerate}[label=(\roman*)]
\item $0 \leq S(\rho) \leq \log d$, and $S(\rho) = \log d$ is achieved on the maximally mixed state $\rho = I/d$.
\item $S(\rho) = 0$ if and only if $\rho$ is pure.
\item Let $\ket{\psi} \in \Hil_{AB}$ be a pure state and $\rho_A = \Tr_B \ketbra{\psi}{\psi}$, $\rho_B = \Tr_A \ketbra{\psi}{\psi}$. Then $S(\rho_A) = S(\rho_B)$.
Furthermore, $S(\rho_A) = S(\rho_B) = 0$ if and only if $\ket{\psi} = \ket{\phi}_A \otimes \ket{\chi}_B$, i.e. $\ket{\psi}$ is \emph{unentangled} across $A$ and $B$.
\item $S(U \rho U^\dagger) = S(\rho)$ for any unitary operator $U$.
\end{enumerate}
\end{prop}

Properties $(i)$ and $(ii)$ tell us that $S(\rho)$ is a measure of purity.
What's more, it gives us a sense of how impure the state is.
Larger values of $S(\rho)$ reflect a larger lack of knowledge about the state $\rho$ if we interpret $\rho$ as a statistical ensemble of its pure eigenbasis states, and the maximum value is achieved on the maximally mixed state.

Property $(iii)$ tells us that entropy is a measure of entanglement in a bipartite system.
For this reason, in a bipartite setting (where the Hilbert space consists of two factors), $S(\rho_A)$ and $S(\rho_B)$ are often called \emph{entanglement entropies}.
Also notice that property $(iv)$ implies that entanglement entropy cannot be changed by acting on a single factor at a time---in order to create entanglement, one must act nonlocally.
When the total state on $\Hil_{AB}$ is pure, there is a precise sense in which entanglement entropy is the \emph{unique} measure that quantifies bipartite entanglement.
Quantifying entanglement when the state on $\Hil_{AB}$ is mixed is a more subtle question (also note that in this case, $S(\rho_A)$ and $S(\rho_B)$ need not be equal).
Section 10.4 of \cite{Preskill:2016htv} is a good point from which to jump into this discussion.

Let's sketch the proof of these properties:\medskip

\noindent {\it Proof sketch of $(i)$:} Working in the eigenbasis of $\rho$, our task is to extremize $S(\rho) \equiv S(p_1, \dots, p_d) = - \sum_i p_i \log p_i$ subject to $0 \leq p_i \leq 1$ and $\sum_i p_i = 1$.
Let $p_d = 1- \sum_{i=1}^{d-1} p_i$ to take care of the latter constraint.
Then, for $1 \leq a \leq d-1$, we have
\begin{equation}
\frac{\partial S}{\partial p_a} = - \log p_a + \log\left(1 - \sum_{i=1}^{d-1} p_i \right) = - \log p_a + \log p_d .
\end{equation}
For there to be a critical point, and hence for $\partial S/\partial p_a$ to vanish, it must be that $p_d = p_a$ for all $1 \leq a \leq d-1$, which is only possible if $p_a = 1/d$ for all $1 \leq a \leq d$.
It's then straightforward to check that this is a maximum, and thus $S(I/d) = \log d$.

~

\noindent {\it Proof sketch of $(ii)$:} Since there was only one critical point of $S(p_1, \dots, p_d)$ and it was a maximum, the minimum must occur on an edge of the domain $0 \leq p_i \leq 1$.
Indeed, at any given edge point where a single $p_i = 1$ and all others vanish, it follows that $S(\rho) = 0$, and this is precisely the case where $\rho$ is pure.

~

\noindent {\it Proof sketch of $(iii)$:} This follows from the Schmidt decomposition (see \Sec{sec:exercises}, Exercise 1).
Given a pure state $\ket{\psi}_{AB}$, there exist orthonormal bases of $\Hil_A$ and $\Hil_B$, $\{\ket{\alpha_i}_A\}_{i=1}^{d_A}$ and $\{\ket{\beta_i}_B\}_{i=1}^{d_B}$, and coefficients $\psi_i$ (some of which could be zero) such that
\begin{equation}
\ket{\psi}_{AB} = \sum_{i=1}^{\min\{d_A,d_B\}} \psi_i \ket{\alpha_i}_A \ket{\beta_i}_B .
\end{equation}
In these bases, $\rho_A$ and $\rho_B$ are both diagonal, and they have the same eigenvalues, $|\psi_i|^2$.
Therefore, it follows that $S(\rho_A) = S(\rho_B)$.
If $\ket{\psi}_{AB} = \ket{\phi}_A \ket{\chi}_B$, then $\rho_A = \ketbra{\phi}{\phi}$ and $\rho_B = \ketbra{\chi}{\chi}$ are both pure, and so their entropies vanish.
Conversely, if $S(\rho_A) = S(\rho_B) = 0$, then $\rho_A$ and $\rho_B$ are both pure states, and so we may write $\rho_A = \ketbra{\phi}{\phi}$ and $\rho_B = \ketbra{\chi}{\chi}$ for some states $\ket{\phi}_A$ and $\ket{\chi}_B$.
Since $\ket{\psi}_{AB}$ is pure by assumption, the total state (already in Schmidt form) must be $\ket{\psi}_{AB} = \ket{\phi}_A \ket{\chi}_B$.

~

\noindent {\it Proof sketch of $(iv)$:} Conjugating a state $\rho$ by a unitary operator does not change its eigenvalues, and so $S(\rho)$ is unchanged.

\eop

Along with these elementary properties, the Von Neumann entropies of reduced states obey many inequalities.
Some of the most important ones are as follows.
\begin{prop} \label{prop:Sineqs}
Some Von Neumann entropy inequalities
\begin{enumerate}[label=(\roman*)]
\item Subadditivity: $S(\rho_{AB}) \leq S(\rho_A) + S(\rho_B)$
\item Araki-Lieb: $|S(\rho_A) - S(\rho_B)| \leq S(\rho_{AB})$
\item Strong Subadditivity: $S(\rho_{AB}) + S(\rho_{BC}) \geq S(\rho_{ABC}) + S(\rho_B)$
\end{enumerate}
\end{prop}
We will not prove these inequalities here, but their proofs may be found in any relatively complete textbook on quantum information (e.g. \cite{nielsen_chuang_2010}).
For brevity, we also often equivalently write $S(\rho_A) \equiv S(A)$.
So, for example, strong subadditivity can be written as $S(AB) + S(BC) \geq S(ABC) + S(B)$.

\subsection{Relative entropy}
\label{subsec:relent}

Having defined Von Neumann entropy, there are many other useful entropic quantities that can be defined and interpreted.
For our purposes, we will need to make extensive use of relative entropy.

\begin{defn}
Let $\rho, \sigma \in \mathcal{S}(\Hil)$.
The \emph{relative entropy} of $\rho$ and $\sigma$ is
\begin{equation}
\rent{\rho}{\sigma} = \Tr(\rho \log \rho) - \Tr(\rho \log \sigma).
\end{equation}
\end{defn}
\Note Relative entropy is only well-defined if the kernel of $\sigma$ is contained in the kernel of $\rho$, denoted $\mrm{ker}~\sigma \subseteq \mrm{ker}~\rho$, or equivalently if the support of $\rho$ is contained in the support of $\sigma$, denoted $\mrm{supp}~\rho \subseteq \mrm{supp}~\sigma$.
In other words, any eigenvector of $\sigma$ with eigenvalue zero must also be an eigenvector of $\rho$ with eigenvalue zero.
This is enough to ensure that $\Tr(\rho \log \sigma)$ is finite.

~

Relative entropy has two key properties that make it a particularly useful quantity.
First, relative entropy is a positive quantity:
\begin{equation}
\rent{\rho}{\sigma} \geq 0 \qquad \text{with equality if and only if} \; \rho = \sigma
\end{equation}
(Exercise~3 in \Sec{sec:exercises} gives a guided derivation of this property.)
We will come back to this property in the next section.

Second, relative entropy obeys an inequality known as \emph{Pinsker's inequality}\footnote{For a proof, see \cite[Chap.~10.8]{Wilde:2011npi}}:
\begin{equation} \label{eq:Pinsker}
\rent{\rho}{\sigma} \geq \frac{1}{2 \log 2} \Vert \rho - \sigma \Vert_1^2
\end{equation}
The one-norm, or trace norm of an operator is defined as
\begin{equation}
\Vert \Oh \Vert_1 = \Tr \sqrt{\Oh^\dagger \Oh}.
\end{equation}
In particular, $\Vert \rho - \sigma \Vert_1$ is a good measure of the \emph{distinguishability} of two states $\rho$ and $\sigma$.
In other words, the smaller the value of $\Vert \rho - \sigma \Vert_1$, then the harder it is to tell the states $\rho$ and $\sigma$ apart using any measurement protocol that you could possibly invent.
(Exercise~2 in \Sec{sec:exercises} makes this explanation precise.)
Pinsker's inequality therefore says that the relative entropy of two states is an upper bound on their distinguishability, and this will play an important role in the holographic application that we will discuss in \Sec{sec:holo}.

\subsection{Application: the black hole information problem}

The small number of basics that we covered in this section already give us enough vocabulary to start asking information-theoretic questions in other areas of physics.
For instance, we can now take up the celebrated black hole information problem \cite{Mathur:2009hf,Polchinski:2016hrw,Harlow:2014yka}, provided that you are willing to take a few facts about black holes and quantum field theory on curved space-time as given.

The earliest version of the black hole information problem is arguably a problem of thermodynamics from the early days of black holes in classical general relativity.
As people realized that black holes---space-time regions whose curvature is such that no object on a causal trajectory can leave the region---were robust predictions of general relativity, they also realized that the following thermodynamic problem had to be taken seriously.
If truly nothing escapes a black hole, then a black hole is a zero-temperature object.
It cannot give off any heat!
This also makes a black hole an entropy sink.
By tossing entropic objects into a black hole, it would seem that you could reduce the total entropy of the universe, in violation of the second law of thermodynamics.

In hindsight, this early black hole ``entropy problem'' is not too hard to patch up.
Owing to initial work on black hole thermodynamics \cite{bardeen1973}, as well as the seminal work of Hawking and Bekenstein \cite{Hawking:1974sw,Bekenstein:1980jp}, we now realize that black holes are indeed well-behaved classical thermodynamic objects.
A black hole has a temperature that depends on its mass, and an entropy that is proportional to the surface area, $A$, of the black hole's event horizon (roughly, the ``point of no return'' from the black hole):
\begin{equation}
S = \frac{A}{4 G_N}
\end{equation}
This formula is known as the Bekenstein-Hawking entropy, $G_N$ is Newton's constant, and we are working in units where $c = \hbar = k_B = 1$.
In particular, tossing an object into a black hole increases its surface area, which hence increases the black hole's entropy, and Bekenstein argued that this increase in black hole entropy would always be enough to preserve the second law of thermodynamics.

While this is a nice resolution from the perspective of classical thermodynamics, the quantum story is quite different.
Hawking argued, based on principles of quantum field theory in curved space-time, that a black hole should radiate particles at a specific temperature.
While this is compelling evidence that black holes obey the laws of thermodynamics, the calculation also comes with the awkward conclusion that the radiation that leaves the black hole is in a \emph{mixed} state.
This is problematic, because nothing in principle prevents us from making a black hole out of matter that is initially in a pure state.
If we let this black hole emit radiation and slowly evaporate away, we are left with a collection of radiation that is in a mixed state at the end of the day.
In other words, it would seem that the formation and subsequent evaporation of a black hole is not a \emph{unitary} process.

Of course, non-unitarity in and of itself is not a problem for quantum mechanics.
When a system is \emph{open}, meaning that it is allowed to exchange information with other degrees of freedom, then generically its evolution will be non-unitary and states that are initially pure can end up mixed.
In fact, we will look at such non-unitary evolution extensively in the next section.
The problem occurs when the system is \emph{closed}.
In this case, when we have truly accounted for all degrees of freedom, quantum evolution should be unitary, so that information does not dissipate away.
Suffice it to say that bad things happen if a closed system evolves non-unitarily, like non-conservation of probabilities.
To Hawking's dismay, his black hole evaporation calculation applies to closed systems.

For a long time, it was believed that subtle corrections to Hawking's calculation would solve the problem---that the radiation that comes out of a black hole is actually in a complicated pure state that only appears thermal on coarse scales.
However, Mathur sharpened the problem in a way that challenges this expectation \cite{Mathur:2009hf}.
Almheiri, Marolf, Polchinski, Stanford, and Sully (collectively referred to as ``AMPSS'') subsequently streamlined the argument\footnote{Many of AMPSS' refinements specifically aimed to rebut a proposal called black hole complementarity \cite{Susskind:1993if}.} by proposing four postulates, each of which seems very reasonable based on what we know about black holes and quantum mechanics:
\begin{enumerate}
\item \emph{Unitarity} -- The formation and evaporation of a black hole is a unitary quantum mechanical process.
\item \emph{Local Effective Field Theory} -- Outside of the horizon of a black hole, physics is well-described by an effective local quantum field theory.
\item \emph{Quantum Black Holes} -- Black holes are themselves quantum mechanical systems with a discrete spectrum of states.
\item \emph{No Drama} -- For a large enough black hole, such that the local curvature at the horizon is very small, nothing special happens to an observer who falls across the horizon into the black hole.
\end{enumerate}
AMPSS then concluded that these postulates cannot all be mutually consistent \cite{Almheiri:2012rt,Almheiri:2013hfa}.

\begin{figure}[ht]
\centering
\includegraphics[width=\textwidth]{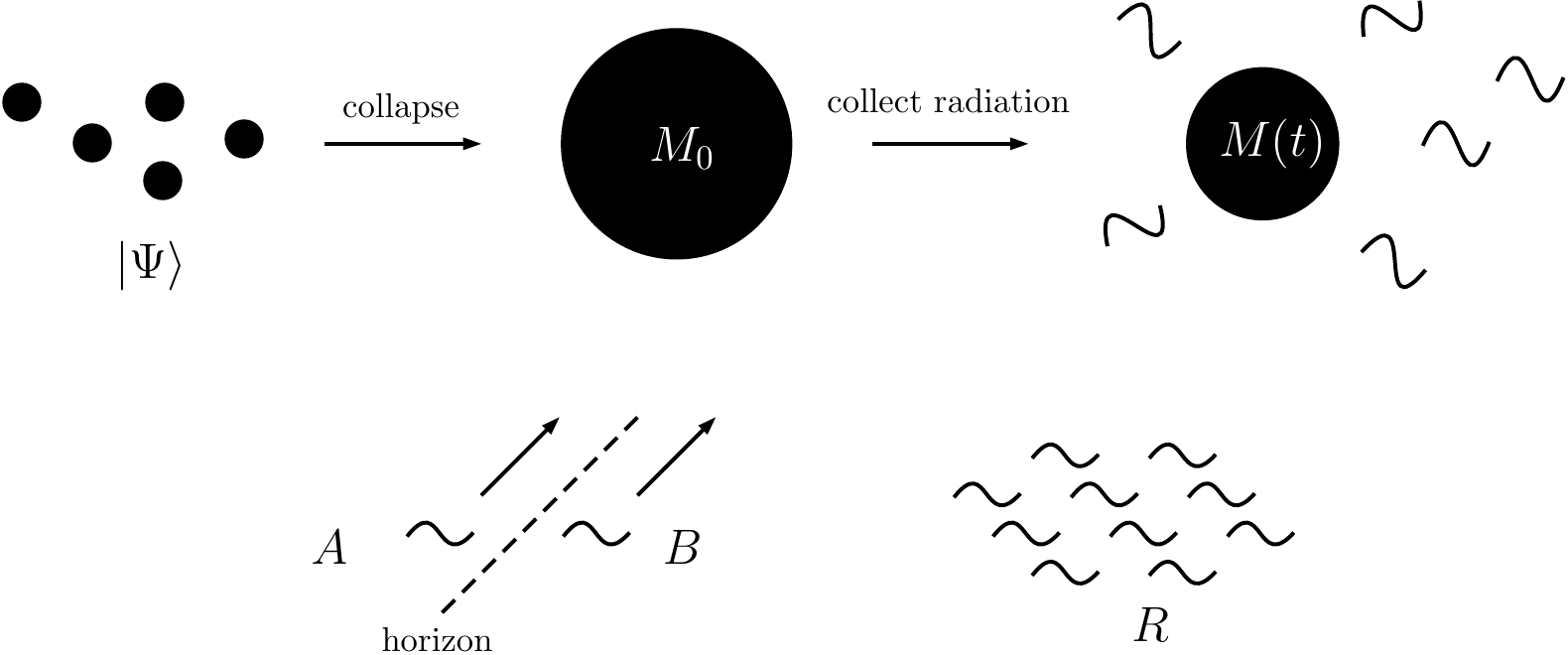}
\caption{The Mathur/AMPSS thought experiment.}
\label{fig:AMPSS}
\end{figure}

Here is a semi-rigorous version of Mathur's argument as rendered by AMPSS, which is illustrated in \Fig{fig:AMPSS}.
Suppose that we begin with a collection of matter that is in some pure state and we collapse it into a black hole of mass $M_0$.
This black hole starts radiating, and we collect all of the radiation that it emits until some time when the mass of the black hole is substantially less than $M_0/2$.
Let $\rho_R$ be the state of the radiation that we have collected.
Consider a particular mode of the radiation---in other words, roughly, a wave-packet of radiation---just outside of the black hole horizon and that is leaving the black hole, and denote its state by $\rho_B$.
From quantum field theoretic arguments, this mode will have a partner mode just inside of the horizon, whose state we denote by $\rho_A$.
Moreover, according to postulates (2) and (4), the joint state of $A$ and $B$ is entangled and pure, meaning that
\begin{equation*}
(i) \qquad S(A) = S(B) \neq 0, \quad S(AB) = 0.
\end{equation*}
Subadditivity of entanglement entropy (Prop.~\ref{prop:Sineqs}-$(i)$) has the saturation property that $S(XY) = S(X) + S(Y)$ if and only if $\rho_{XY} = \rho_X \otimes \rho_{Y}$.
Since $\rho_{AB}$ is pure, it follows that $\rho_{ABR} = \rho_{AB} \otimes \rho_R$, and so
\begin{equation*}
(ii) \qquad S(ABR) = S(R).
\end{equation*}

Next, postulate (1) implies that
\begin{equation*}
(iii) \qquad S(BR) < S(R).
\end{equation*}
This is the mathematical statement that once the black hole has lost roughly half of its initial mass to evaporation\footnote{More precisely, past the \emph{Page time}, at which point the black hole's horizon area reaches half its initial value.}, any quantum of radiation that subsequently leaves the black hole should purify the radiation that came out earlier if black hole evaporation is unitary.
(At the end of unitary evaporation, we must have that $S(R) = 0$, since $R$ is all that is left.)
Finally, we also have strong subadditivity (Prop.~\ref{prop:Sineqs}-$(iii)$) among the $A$, $B$, and $R$ subsystems:
\begin{equation*}
(iv) \qquad S(AB) + S(BR) \geq S(ABR) + S(B)
\end{equation*}

Putting it all together, we find the following:
\begin{align*}
S(R) + S(B) &= S(ABR) + S(B) & \text{using}~(ii) \\
&\leq S(AB) + S(BR) & \text{using}~(iv) \\
&< S(AB) + S(R) & \text{using}~(iii) \\
&= S(R) & \text{using}~(i)
\end{align*}
Since $S(B) \neq 0$, we have therefore arrived at a contradiction!

AMPSS' conclusion was that one of their four postulates has to be modified.
How palatable the ensuing consequences are is up to you to reason through.
\begin{enumerate}
\item If we drop unitarity, then black holes destroy quantum information \cite{Unruh:2017uaw}.
\item One way to modify local effective field theory is to delete the word ``local'' and allow for small amounts of nonlocality \cite{Giddings:2012gc}, although such an approach is not without its rebuttals, e.g. \cite[Sec.~8]{Raju:2020smc}.
Holographic resolutions of the black hole information problem (and AdS/CFT itself) are also nonlocal in the sense that degrees of freedom are replicated in both the bulk space-time and its boundary (see \Sec{sec:holo}).
\item One way to evade AMPSS' argument is if black holes never finish evaporating and instead leave behind a small and extremely entropic remnant \cite{Chen:2014jwq}.
Or, perhaps black holes are just not described by quantum mechanics.
\item If $A$ and $B$ are not in a pure entangled state such that $S(AB) \neq 0$, then it's possible to evade the contradiction.
However, such states have large local energy densities.
In our setting, it would be as if there was a firewall waiting just behind the horizon that an infalling observer would hit as they entered the black hole.
As AMPSS pointed out, the result is considerable drama for the observer.
\end{enumerate}
The references given above are by no means a complete account of the literature and only represent a handful of the big ideas in their corresponding directions.
The second AMPSS paper \cite{Almheiri:2013hfa} is a traditional place to start looking for more literature if you want to learn more about different approaches to the black hole information problem.
Refs.~\cite{Mathur:2009hf,Harlow:2014yka,Polchinski:2016hrw} are accessible and pedagogical reviews, and \Ref{Raju:2020smc} gives a particularly thorough and modern review of the subject.

Recent attempts at resolving the black hole information problem have focused on black holes in AdS/CFT (for a review, see \cite{Almheiri:2020cfm}).
In this setting at least, it seems that unitarity is maintained by having the Hawking radiation encode the interior of the black hole so that a violation of strong subadditivity is avoided.
Morally, these resolutions are a relaxation of locality, since distant Hawking radiation encodes a faraway region inside the black hole, but this nonlocality is no more drastic than holography itself, in which distant degrees of freedom at the boundary of a space-time encode physics deep inside.
Whether and how this reasoning extends to more general black holes is a topic of current research.

\section{Quantum channels}
\label{sec:channels}

We largely focused on properties of states in the last section.
In this section, we will study how quantum states evolve.
When we first learn about quantum mechanics, we learn about unitary evolution according to the Schr{\"o}dinger equation.
But, as you may already be aware of, much more general yet physically reasonable quantum evolution is possible.
Such evolution is described by quantum channels.

\subsection{Definition and properties}

Informally, a \emph{quantum channel} is a map that sends states to states.
\begin{eg}
Unitary evolution is a quantum channel. Let $\rho \in \mathcal{S}(\Hil)$ and $U \in \mathcal{L}(\Hil)$ be a unitary operator.
The map
\begin{equation}
\begin{aligned}
\mathcal{N}_U : ~ &\mathcal{S}(\Hil) \rightarrow \mathcal{S}(\Hil) \\
& \rho \mapsto U \rho U^\dagger
\end{aligned}
\end{equation}
is a quantum channel.
\eop
\end{eg}

\begin{eg} \label{eg:nonu}
A channel can also describe non-unitary evolution. Let Let $\rho_A \in \mathcal{S}(\Hil_A)$, $\ket{0}_B \in \Hil_B$ be some fixed state, and let $U_{AB} \in \mathcal{L}(\Hil_{AB})$ be a unitary operator.
The map
\begin{equation}
\begin{aligned}
\mathcal{N} : ~ &\mathcal{S}(\Hil_A) \rightarrow \mathcal{S}(\Hil_A) \\
& \rho_A \mapsto \Tr_B\left[ U_{AB} \left(\rho_A \otimes \ketbra{0}{0}_B \right) U_{AB}^\dagger \right]
\end{aligned}
\end{equation}
is a quantum channel.
For generic choices of $U_{AB}$, $\mathcal{N}(\rho_A)$ will not in general be pure even if $\rho_A$ is pure.
\eop
\end{eg}

Let's now be a bit more systematic.
Let $\mathcal{N} : \mathcal{L}(\Hil_A) \rightarrow \mathcal{L}(\Hil_B)$ be a map from linear operators on $\Hil_A$ to linear operators on $\Hil_B$.
If we want $\mathcal{N}$ to map states to states, what is the minimal set of properties that should it have?

~

\noindent 1. $\mathcal{N}$ should be \emph{trace-preserving}, i.e.,
\begin{equation}
\Tr_B[\mathcal{N}(\Oh)] = \Tr_A[\Oh].
\end{equation}
This will ensure that the image of a density operator will still have unit trace.

~

\noindent 2. $\mathcal{N}$ should be \emph{linear}, i.e.,
\begin{equation}
\mathcal{N}(\lambda_1 \Oh_1 + \lambda_2 \Oh_2) = \lambda_1 \mathcal{N}(\Oh_1) + \lambda_2 \mathcal{N}(\Oh_2)
\end{equation}
for all $\lambda_1, \lambda_2 \in \mathbb{C}$.
This is reasonable to require so that the ensemble interpretation of density operators continues to hold.
Explicitly, suppose that we decompose a density operator as a probabilistic ensemble,
\begin{equation}
\rho = \sum_i p_i \rho_i,
\end{equation}
for a collection of density operators $\rho_i$ and probabilities $p_i \in [0,1]$ such that $\sum_i p_i = 1$.
The interpretation of such an ensemble is that the state $\rho$ describes a configuration where the state $\rho_i$ is prepared with probability $p_i$.
If we send $\rho$ through the channel $\mathcal{N}$, it should then be that the state $\mathcal{N}(\rho_i)$ occurs with probability $p_i$, i.e., $\mathcal{N}(\rho) = \sum_i p_i \mathcal{N}(\rho_i)$.

Still, it's fun to ask what happens if a map between states is nonlinear.
The next example demonstrates a specific strange occurrence.

\begin{eg}
Consider the map whose action on a qubit state $\rho \in \mathcal{S}(\Hil)$, $\Hil = \mrm{span} \{\ket{0},\ket{1}\}$, is given by
\begin{equation}
\mathcal{E}(\rho) = e^{i\pi X \Tr[X \rho]} ~ \rho ~ e^{-i \pi X \Tr[X \rho]},
\end{equation}
where $X$ is the Pauli $x$ operator (i.e. $X\ket{0} = \ket{1}$ and $X \ket{1} = \ket{0}$).
This map is trace-preserving (as can be seen using the cyclic property of the trace), but it is clearly not a linear map.
In a first scenario, suppose that we prepare a state $\rho_1 = \tfrac{1}{2} \ketbra{0}{0} + \tfrac{1}{2} \ketbra{1}{1}$.
Since $\Tr[X \rho_1] = 0$, it follows that $\mathcal{E}(\rho_1) = \rho_1$.
In a second scenario, however, suppose that we first prepare $\rho_1$ and then perform an operation such that if the state $\ket{1}$ is prepared, it gets rotated to the state $\ket{+} = \tfrac{1}{\sqrt{2}}(\ket{0} + \ket{1})$, resulting in a state $\rho_2 = \tfrac{1}{2} \ketbra{0}{0} + \tfrac{1}{2}\ketbra{+}{+}$.
Since $\Tr[X \rho_2] = \tfrac{1}{2}$, it follows that $\mathcal{E}(\rho_2) = X \rho_2 X = \tfrac{1}{2} \ketbra{1}{1} + \tfrac{1}{2} \ketbra{+}{+}$.

This is very strange evolution in light of the ensemble interpretation of density operators.
Comparing the two scenarios, we see that the state $\ketbra{0}{0}$, which is prepared with probability $\tfrac{1}{2}$ in each case, evolves differently depending on how we \emph{would have} prepared the other state had we not prepared $\ketbra{0}{0}$.
In other words, $\mathcal{E}$ describes evolution that depends on possibilities that are not actually realized.
\eop
\end{eg}

\noindent Since density operators describe probabilities, $\mathcal{N}$ itself should certainly be \emph{positive}, i.e., if $\Oh$ is positive semi-definite, then $\mathcal{N}(\Oh)$ should also be positive semi-definite.
This is the strict minimum needed to ensure that the image of a density operator is positive semi-definite, but we will actually require something a bit stronger:

~

\noindent 3. $\mathcal{N}$ should be \emph{completely positive}. Given any other auxiliary Hilbert space $\Hil_R$, we require that the map
\begin{equation}
\mrm{id}_R \otimes \mathcal{N} : \mathcal{L}(\Hil_R \otimes \Hil_A) \rightarrow \mathcal{L}(\Hil_R \otimes \Hil_B)
\end{equation}
is positive, where $\mrm{id}_R$ is the identity map on $\mathcal{L}(\Hil_R)$.

This requirement should seem fairly innocuous, and it's certainly reasonable on physical grounds.
If $A$ is the part of the universe under consideration and $R$ is some other part, or even the rest of the universe itself, then evolving $A$ with $\mathcal{N}$ and doing nothing to the rest of the universe should map a state of the universe to a state of the universe.
It turns out that complete positivity will let us prove a powerful result about channels (Thm.~\ref{thm:choi-kraus} below).
Before doing this, let's see an example of a map that is positive, but not completely positive.

\begin{eg}
Let $\Hil = \Span \{\ket{i} \}_{i=1}^d$.
The transpose map
\begin{equation}
{}^T : \ketbra{i}{j} \mapsto \ketbra{j}{i}
\end{equation}
is a positive map.
If $\Oh = \sum_{ij} \Oh_{ij} \ketbra{i}{j}$ is positive semi-definite, then for any $\ket{\psi} = \sum_i \psi_i \ket{i}$, we have that
\begin{equation}
\bra{\psi} \Oh^T \ket{\psi} = \sum_{ij} \psi_i^* (\Oh^T)_{ij} \psi_j = \sum_{ij} \psi_j \Oh_{ji} \psi_i^* = \bra{\psi^*} \Oh \ket{\psi^*} \geq 0,
\end{equation}
where $\ket{\psi^*}$ denotes the state $\sum_i \psi_i^* \ket{i}$.
However, let $\Hil \equiv \Hil_A$, and suppose that we augment the Hilbert space with $\Hil_R \cong \Hil_A$.
Define the (unnormalized) maximally entangled state
\begin{equation} \label{eq:maxent}
\ket{\Gamma}_{RA} = \sum_i \ket{i}_R \ket{i}_A
\end{equation}
and consider the action of $\mrm{id}_R \otimes {}^{T} $ on $\ketbra{\Gamma}{\Gamma}_{RA}$:
\begin{align*}
(\mrm{id}_R \otimes {}^{T})(\ketbra{\Gamma}{\Gamma}_{RA}) &= (\mrm{id}_R \otimes {}^{T})\left( \sum_{ij} \ketbra{i}{j}_R \otimes \ketbra{i}{j}_A \right) \\
&= \sum_{ij} \ketbra{i}{j}_R \otimes \ketbra{j}{i}_A \\
&\equiv \mrm{SWAP}_{RA}
\end{align*}
$\ketbra{\Gamma}{\Gamma}_{RA}$ therefore maps to the SWAP operator, which interchanges the state on $A$ with the state on $R$.
However, $(\mrm{SWAP})^2 = I$, which means that the eigenvalues of SWAP are $\pm 1$.
Since SWAP has negative eigenvalues, it is not a positive semi-definite operator.
\eop
\end{eg}

We can now give a formal definition of a quantum channel:

\begin{defn}
A \emph{quantum channel} is a map $\mathcal{N} : \mathcal{L}(\Hil_A) \rightarrow \mathcal{L}(\Hil_B)$ that is linear, trace-preserving, and completely positive.
\end{defn}

\subsection{The operator-sum representation}

A further motivation for requiring complete positivity is that it lets us prove the following theorem, which is a powerful characterization of the general structure of quantum channels.
We will first state the theorem, look at a simple example, and then go on to prove the theorem.
The proof is a mix of the proofs given by Refs.~\cite{preskill} and \cite{Wilde:2011npi}, and it includes a few of my own touches.
Following Wilde's notation, we will sometimes add a subscript to a channel to indicate its domain and range.
\begin{thm}[Choi-Kraus] \label{thm:choi-kraus}
A linear map $\mathcal{N}_{A \rightarrow B} : \mathcal{L}(\Hil_A) \rightarrow \mathcal{L}(\Hil_B)$ is completely positive and trace-preserving (CPTP) if and only if
\begin{equation} \label{eq:opsum}
\mathcal{N}_{A \rightarrow B}(X_A) = \sum_{\ell = 1}^d M_\ell X_A M_\ell^\dagger
\end{equation}
for all $X_A \in \mathcal{L}(\Hil_A)$, where the $M_\ell \in \mathcal{L}(\Hil_A, \Hil_B)$ are linear maps from $\Hil_A$ to $\Hil_B$ satisfying
\begin{equation}
\sum_{\ell=1}^d M_\ell^\dagger M_\ell = I_A
\end{equation}
and that may be chosen such that $d \leq d_A d_B$.
\end{thm}
\Note \Eq{eq:opsum} is the \emph{operator-sum} representation of $\mathcal{N}_{A \rightarrow B}$ and
the operators $M_\ell$ are called \emph{Kraus operators}.
The Kraus operators for a given channel are not unique, but we will come back to this point in \Sec{sec:isometricdilation}.

\begin{eg}
Reconsider the channel from Ex.~\ref{eg:nonu}:
\begin{align*}
\mathcal{N}(\rho_A) &= \Tr_B \left[ U_{AB} \left(\rho_A \otimes \ketbra{0}{0}_B \right) U_{AB}^\dagger \right] \\
&= \sum_{j=1}^{d_B} {}_B\bra{j} U_{AB} \ket{0}_B \, (\rho_A) \, {}_B\bra{0} U_{AB}^\dagger \ket{j}_B \\
&\equiv \sum_{j=1}^{d_B} M_j \rho_A M_j^\dagger
\end{align*}
The operators $M_j$ are linear, and we can check the completeness relation:
\begin{align*}
\sum_j M_j^\dagger M_j &= \sum_j  {}_B\bra{0} U_{AB}^\dagger \ketbra{j}{j} U_{AB} \ket{0}_B \\
&= {}_B\bra{0} U_{AB}^\dagger \left( \sum_j \ketbra{j}{j} \right) U_{AB} \ket{0}_B \\
&= {}_B\bra{0} U_{AB}^\dagger U_{AB} \ket{0}_B \\
&= {}_B\bra{0} I_{AB} \ket{0}_B \\
&= I_A
\end{align*}
We have therefore exhibited an operator-sum decomposition of $\mathcal{N}$ and a set of Kraus operators.
\eop
\end{eg}

\noindent {\bf Proof (Choi-Kraus Theorem):} First we prove the forward direction.
Suppose that the action of $\mathcal{N}_{A \rightarrow B}$ is given by \Eq{eq:opsum}.
This action clearly defines a linear map.
To establish complete positivity, consider the action of $\mrm{id}_R \otimes \mathcal{N}_{A \rightarrow B}$ on a positive semi-definite operator $X_{RA} \in \mathcal{L}(\Hil_{RA})$:
\begin{equation}
(\mrm{id}_R \otimes \mathcal{N}_{A \rightarrow B})(X_{RA}) = \sum_\ell (I_R \otimes M_\ell) X_{RA} (I_R \otimes M_\ell^\dagger)
\end{equation}
Given any state $\ket{\psi}_{RB} \in \Hil_{RB}$, if we define the state $\ket{\tilde \psi_\ell}_{RA} = (I_R \otimes M_\ell^\dagger)\ket{\psi}_{RB}$, for each $\ell$ we can write
\begin{equation}
{}_{RB}\bra{\psi} (I_R \otimes M_\ell) X_{RA} (I_R \otimes M_\ell^\dagger) \ket{\psi}_{RB} = {}_{RA}\bra{\tilde \psi_\ell} X_{RA} \ket{\tilde \psi_\ell}_{RA} \geq 0.
\end{equation}
Therefore, ${}_{RB}\bra{\psi}(\mrm{id}_R \otimes \mathcal{N}_{A \rightarrow B})(X_{RA})\ket{\psi}_{RB} \geq 0$, and so $\mathcal{N}_{A \rightarrow B}$ is completely positive.
To check that $\mathcal{N}_{A \rightarrow B}$ is trace-preserving, we just calculate.
Let $X_A \in \mathcal{L}(\Hil_A)$:
\begin{align*}
\Tr_B \left[ \mathcal{N}_{A \rightarrow B}(X_A) \right] &= \Tr_B \left[ \sum_\ell M_\ell X_A M_\ell^\dagger \right] \\
&= \Tr_A \left[ \sum_\ell M_\ell^\dagger M_\ell X_A  \right] \\
&= \Tr_A [X_A]
\end{align*}
Checking that the cyclic property of the trace still holds for the partial traces above (i.e., going from the first to the second line) is the short Exercise~4 in \Sec{sec:exercises}.

Next we prove the reverse direction.
Suppose that $\mathcal{N}_{A \rightarrow B} : \mathcal{L}(\Hil_A) \rightarrow \mathcal{L}(\Hil_B)$ is a linear, CPTP map.
We must show that it has an operator-sum representation.
First, let us make a brief digression to introduce a useful tool:
\begin{defn} \label{defn:Choi}
Let $\Hil_R \cong \Hil_A$ and recall the unnormalized maximally entangled state $\ket{\Gamma}_{RA}$ defined in \Eq{eq:maxent}.
The \emph{Choi operator} is the operator
\begin{equation}
(\mrm{id}_R \otimes \mathcal{N}_{A \rightarrow B})(\ketbra{\Gamma}{\Gamma}_{RA}) = \sum_{i,j=1}^{d_A} \ketbra{i}{j}_R \otimes \mathcal{N}_{A \rightarrow B}(\ketbra{i}{j}_A).
\end{equation}
\end{defn}
Next, we make two observations.
First, since $\mathcal{N}_{A \rightarrow B}$ is completely positive, the Choi operator is itself a (non-normalized) state.
We can therefore diagonalize it and write
\begin{equation} \label{eq:channel2state}
(\mrm{id}_R \otimes \mathcal{N}_{A \rightarrow B})(\ketbra{\Gamma}{\Gamma}_{RA}) = \sum_{\ell = 1}^d \ketbra{\phi_\ell}{\phi_\ell}_{RB}
\end{equation}
for some (non-normalized) non-zero vectors $\{ \ket{\phi_\ell}_{RB} \}_{\ell = 1}^{d}$, where $d \leq d_R d_B = d_A d_B$.
Second, given any vector $\ket{\psi}_A \in \Hil_A$, we can write
\begin{equation}
\ket{\psi}_A = \sum_{i=1}^{d_A} \psi_i \, \ket{i}_A = \sum_{i=1}^{d_A} \psi_i ({}_R \braket{i}{\Gamma}_{RA} ) = {}_R\braket{\psi^*}{\Gamma}_{RA}.
\end{equation}
Putting these two observations together, for $\ket{\psi}_A, \ket{\chi}_A \in \Hil_A$, we find the following:
\begin{align*}
\mathcal{N}_{A \rightarrow B}(\ketbra{\psi}{\chi}_A) &= \mathcal{N}_{A \rightarrow B} \left( {}_R\braket{\psi^*}{\Gamma} \braket{\Gamma}{\chi^*}_R \right) \\
&= {}_R\bra{\psi^*} (\mrm{id}_R \otimes \mathcal{N}_{A \rightarrow B})(\ketbra{\Gamma}{\Gamma}_{RA}) \ket{\chi^*}_R \\
&= \sum_{\ell = 1}^d {}_R\bra{\psi^*} (\ketbra{\phi_\ell}{\phi_\ell}_{RB} ) \ket{\chi^*}_R
\end{align*}
With this in mind, for each $\ell$, define a linear operator
\begin{equation} \label{eq:Mops}
\begin{aligned}
M_\ell : ~ &\Hil_A \rightarrow \Hil_B \\
&\ket{\psi}_A \rightarrow {}_R\braket{\psi^*}{\phi_\ell}_{RB}
\end{aligned}
\end{equation}
with an adjoint that satisfies
\begin{equation} \label{eq:Madj}
{}_A\bra{\psi} M_\ell^\dagger = (M_\ell \ket{\psi}_A )^\dagger = {}_{RB} \braket{\phi_\ell}{\psi^*}_R .
\end{equation}
We can therefore write
\begin{equation} \label{eq:state2channel}
\mathcal{N}_{A \rightarrow B}(\ketbra{\psi}{\chi}_A) = \sum_{\ell = 1}^d M_\ell \ketbra{\psi}{\chi}_A M_\ell^\dagger.
\end{equation}
Any linear operator $X_A$ can be written as a sum over single-rank operators like $\ketbra{\psi}{\chi}$, and so by linearity, we we have that
\begin{equation}
\mathcal{N}_{A \rightarrow B}(X_A) = \sum_{\ell=1}^d M_\ell X_A M_\ell^\dagger
\end{equation}
for all $X_A \in \mathcal{L}(\Hil_A)$.
The last thing that we have to show is that the $M_\ell$ obey the required completeness relation.
To this end, we exploit the fact that $\mathcal{N}_{A \rightarrow B}$ is trace preserving:
\begin{equation}
\Tr_B[\mathcal{N}_{A \rightarrow B}(\ketbra{i}{j}_A)] = \Tr_A[ \ketbra{i}{j}_A] = \delta_{ij}
\end{equation}
However, according to the operator-sum decomposition that we found,
\begin{align*}
\Tr_B[\mathcal{N}_{A \rightarrow B}(\ketbra{i}{j}_A)] &= \Tr_B \left[ \sum_{\ell=1}^d M_\ell \ketbra{i}{j}_A M_\ell^\dagger  \right] \\
&= \Tr_A \left[ \sum_{\ell=1}^d  M_\ell^\dagger M_\ell   \ketbra{i}{j}_A \right] \\
&= \bra{j} \left( \sum_{\ell=1}^d  M_\ell^\dagger M_\ell \right) \ket{i}_A
\end{align*}
Therefore, it must be that $\sum_{\ell=1}^d M_\ell^\dagger M_\ell = I_A$, which completes the proof of the theorem.

\eop

We were a bit quick about it in the proof above, but it's worth noting that $M_\ell^\dagger$ as defined through \Eq{eq:Madj} is indeed a well-defined map from $\Hil_B$ to $\Hil_A$.
From our definitions, we can write the following:
\begin{align*}
{}_A \bra{\psi} M_\ell^\dagger \ket{\chi}_B &= {}_{RB}\braket{\phi_\ell}{\psi^*}_R \ket{\chi}_B \\
&= {}_R \bra{\psi} {}_B \braket{\chi^*}{\phi^*}_{RB} \\
&= {}_R\bra{\psi} (M^\dagger_\ell \ket{\chi}_B )_R
\end{align*}
Since $\Hil_R \cong \Hil_A$, we can relabel the last lines to define the action of $M_\ell^\dagger$ as
\begin{equation}
M_\ell^\dagger : \ket{\chi}_B \mapsto {}_B\braket{\chi^*}{\phi^*}_{AB}.
\end{equation}

\subsection{Further properties and results}

In the last part of this section, we examine a handful of further properties of channels in light of the Choi-Kraus theorem and its proof.

\subsubsection{Channel-state duality}

The Choi operator (Def.~\ref{defn:Choi}) that we introduced during the proof of Thm.~\ref{thm:choi-kraus} defines a one-to-one correspondence between states and channels that is known as \emph{channel-state duality}, or the \emph{Choi-Jamiolkowski isomorphism}.
The Choi operator itself, via \Eq{eq:channel2state}, associates a state to a given channel $\mathcal{N}_{A\rightarrow B}$ that encodes all of the channel's properties, including its action.
Conversely, given any state on a Hilbert space $\Hil_{RB}$, which we write in diagonal form as $\sum_{\ell=1}^d \ketbra{\phi_\ell}{\phi_\ell}_{RB}$, \Eq{eq:state2channel} and the operators $M_\ell$ defined by \Eq{eq:Mops} together define a channel $\mathcal{N}_{A\rightarrow B}$ from a Hilbert space $\Hil_A \cong \Hil_R$ to $\Hil_B$.
We will not make any further use of channel-state duality, but it's worth being aware of since it's a useful tool in quantum information theory that you will surely encounter again.

\subsubsection{Isometric dilation}
\label{sec:isometricdilation}

An important consequence of the Choi-Kraus theorem is that we can always think of a channel as coming from an isometric operator, called its \emph{isometric dilation}, on a larger Hilbert space.
\begin{prop}
Let $\mathcal{N}_{A \rightarrow B} : \mathcal{L}(\Hil_A) \rightarrow \mathcal{L}(\Hil_B)$ be a CPTP map, and let $\Hil_E$ be an auxiliary Hilbert space such that $\dim \Hil_E \geq d$, where $d$ is as defined in Thm.~\ref{thm:choi-kraus}.
Then, there exists a linear isometry $V : \Hil_A \rightarrow \Hil_B \otimes \Hil_E$ such that
\begin{equation}
\Tr_E[V X_A V^\dagger] = \mathcal{N}_{A \rightarrow B}(X_A)
\end{equation}
for all $X_A \in \mathcal{L}(X_A)$, where $V^\dagger V = I_A$ and $VV^\dagger = \Pi_{BE}$.
The operator $\Pi_{BE}$ denotes the projector onto the image of $\mathcal{L}(\Hil_A)$ under $V$.
\end{prop}

Several comments are in order.
First, an \emph{isometry} is an inner product-preserving map linear map, i.e. $\braket{V\phi}{V\psi} = \braket{\phi}{\psi}$.
Furthermore, it's easy to extend an isometry to a unitary operator by adding extra Hilbert spaces to its domain or range.
For example, let $\Hil_A = \Span \{\ket{0}_A\}$, $\Hil_B = \Span \{\ket{0}_B, \ket{1}_B\}$, and define the isometry $V : \ket{0}_A \mapsto \tfrac{1}{\sqrt{2}} (\ket{0}_B + \ket{1}_B)$.
If we introduce an extra Hilbert space $\Hil_{A'} = \Span \{\ket{1}_{A'}\}$, then it's straightforward to extend $V$ to a unitary operator $U : \Hil_{A} \oplus \Hil_{A'} \rightarrow \Hil_B$ such that the restriction of $U$ to $\Hil_A$ is $V$, i.e. $U|_A = V$.
For instance,
\begin{equation}
\begin{aligned}
U : ~ &\ket{0}_A \mapsto \frac{1}{\sqrt{2}} (\ket{0}_B + \ket{1}_B) \\
&\ket{1}_{A'} \mapsto \frac{1}{\sqrt{2}} (\ket{0}_B - \ket{1}_B)
\end{aligned}
\end{equation}
does the trick.
The take-home message of this discussion is that it is always possible to think of non-unitary evolution in a given Hilbert space as unitary evolution in a larger Hilbert space in which we are ignorant of certain degrees of freedom.

Given that $V^\dagger V = I_A$, it's straightforward to see that $VV^\dagger$ has to be a projector, because
\begin{equation}
(VV^\dagger)(VV^\dagger) = V(V^\dagger V)V^\dagger = V I_A V^\dagger = VV^\dagger.
\end{equation}
The isometry $V$ is itself easy to construct.
Let $\{M_j\}$ be a set of Kraus operators for $\mathcal{N}_{A\rightarrow B}$, and let $\{ \ket{e_j}_E \}$ be an orthonormal basis for $\Hil_E$.
Then, $V$ is given by
\begin{equation}
V = \sum_{j=1}^d M_j \otimes \ket{e_j}_E,
\end{equation}
and its action on a state $\ket{\psi}_A \in \Hil_A$ is
\begin{equation}
V\ket{\psi}_A = \sum_{j=1}^d (M_j \ket{\psi}_A) \otimes \ket{e_j}_E.
\end{equation}
We can check that $V^\dagger V = I_A$:
\begin{align*}
V^\dagger V &= \sum_{i,j} M_i^\dagger M_j \braket{e_i}{e_j}_E \\
&= \sum_i M_i^\dagger M_i \\
&= I_A
\end{align*}
Finally, tracing out $E$ indeed reproduces the action of $\mathcal{N}_{A\rightarrow B}$:
\begin{align*}
\Tr_E[V X_A V^\dagger] &= \Tr_E \left[ \sum_{i,j} M_i X_A M_j^\dagger \otimes \ketbra{e_i}{e_j}_E  \right] \\
&= \sum_{i,j} M_i X_A M_j^\dagger \braket{e_j}{e_i}_E \\
&= \sum_{i} M_i X_A M_i^\dagger \\
&= \mathcal{N}_{A \rightarrow B}(X_A)
\end{align*}

The isometric dilation also gives us a way to easily show that the choice of Kraus operators in an operator-sum decomposition is not unique.
Suppose we perform a unitary change of basis in $\Hil_E$ and write
\begin{equation}
\ket{e_i}_E = \sum_j W_{ij} \ket{\tilde e_j}_E.
\end{equation}
Then $V$ remains an isometric dilation of $\mathcal{N}_{A \rightarrow B}$, but we see that
\begin{equation}
V = \sum_{i} M_i \otimes \sum_j W_{ij} \ket{\tilde e_j}_E = \sum_j \left(\sum_i W_{ij} M_i \right) \otimes \ket{\tilde e_j}_E \equiv \sum_j N_j \otimes \ket{\tilde e_j}_E,
\end{equation}
and so we have found another set of Kraus operators, $\{ N_j \}$.
It turns out that two sets of Kraus operators are always related unitarily in this way if they correspond to the same channel.
For a proof, see \cite{preskill}.

\subsubsection{Monotonicity of relative entropy}

The final topic we will look at in this section is a combined property of relative entropy and channels:

\begin{thm}[Monotonicity of relative entropy] \label{thm:Dmono}
Let $\mathcal{N} : \mathcal{L}(\Hil_A) \rightarrow \mathcal{L}(\Hil_B)$ be a CPTP map.
Then, for all $\rho, \sigma \in \mathcal{S}(\Hil_A)$, it follows that $\rent{\rho}{\sigma} \geq \rent{\mathcal{N}(\rho)}{\mathcal{N}(\sigma)}$.
\end{thm}

The mathematical content of this theorem is that evolution by a channel can never cause the relative entropy between two states to increase.
In light of our discussion from \Sec{subsec:relent}, the physical content of this theorem is that a channel can only degrade states.
At best, a pair of states can only remain as distinguishable as they were before.
For a proof of this theorem, see \cite[Thm.~11.8.1]{Wilde:2011npi}.

\section{Quantum error correction}
\label{sec:QEC}

As we briefly touched on in the introduction, errors are certain to occur whenever we try to implement a quantum computation.
Whether they are due to unwanted interactions with the surrounding environment, faulty implementations of unitary operations, or some other reason, we need a way to protect computations from errors.
This is what we achieve with quantum error correction.\footnote{Quantum error correction is distinct from \emph{fault tolerance}, which is equally crucial for computation, but which we will not cover here. For an introduction, see \cite{Devitt_2013}.}

The basic idea of quantum error correction is to embed a smaller Hilbert space, called the \emph{logical space} or \emph{code subspace}, into a larger Hilbert space, called the \emph{physical space}:
\begin{equation*}
\Hil_\mrm{code} \hookrightarrow \Hil_\mrm{phys}
\end{equation*}
The actual physical degrees of freedom of a quantum computer are described by $\Hil_\mrm{phys}$, but the logical computation that we want to achieve takes place in $\Hil_\mrm{code}$.
A specific embedding is a \emph{quantum error correcting code} (QECC), and for a QECC to be good, it must protect the logical computation from errors that are likely to occur.
More precisely, this means that we must be able to use the extra degrees of freedom afforded by $\Hil_\mrm{phys}$ to monitor the computer's state for errors, and we must able to correct these errors when they occur.
We typically expect that errors tend to be largely uncorrelated and localized in the physical space,\footnote{From a practical standpoint, a QECC is only as good as the extent to which the errors it is designed to correct faithfully model the errors that actually occur.} and so good QECCs tend to encode the logical information \emph{nonlocally} in $\Hil_\mrm{phys}$
What's more, we have to be very clever in how we carry out monitoring and error recovery tasks so as not to disturb the information contained in the computer's computational state!

Rather than dwell further on abstract features, the best way to become familiar with quantum error correction is to see an example of a QECC.
This is what we will do in the first part of this section.
In the second part, we will reformulate quantum error correction in the language of quantum channels, which ties into the previous section and sets us up for the holographic applications discussed in the next.

\subsection{Two quantum error correcting codes}

Before we begin, let us briefly confirm some notation and conventions.
The Hilbert space of a single qubit is spanned by two basis vectors, $\ket{0}$ and $\ket{1}$, that are eigenstates of the Pauli $z$ operator, which we denote by $Z$, i.e.,
\begin{equation}
Z\ket{0} = \ket{0} \qquad Z\ket{1} = -\ket{1}.
\end{equation}
Similarly, the other Pauli operators are denoted by $X$ and $Y$.
We will denote a basis state for $n$ qubits by a binary string,
\begin{equation}
\ket{x_1} \otimes \ket{x_2} \otimes \dots \otimes \ket{x_n} \equiv \ket{x_1 x_2 \dots x_n}, \quad x_i \in \{0,1\} ~\text{for}~ 1\leq i \leq n,
\end{equation}
and we call this basis the \emph{computational basis}.
Finally, we indicate that a single-qubit operator $\Oh$ acts on the $i^\text{th}$ qubit with a subscript, $\Oh_i$, and we usually suppress any identity operators and tensor product symbols.
For example,
\begin{align*}
Z_1 X_3 Z_4X_4 \ket{x_1 x_2 x_3 x_4} &\equiv (Z_1 \otimes I_2 \otimes X_3 \otimes Z_4 X_4) \ket{x_1 x_2 x_3 x_4} \\
&= (Z_1 \ket{x_1}) \otimes \ket{x_2} \otimes (X_3 \ket{x_3}) \otimes (Z_4 X_4 \ket{x_4}).
\end{align*}

\subsubsection{A rudimentary 3-qubit code}

Suppose that we want to design a QECC for a single logical qubit.
For our first attempt, suppose that we have three physical qubits at our disposal and that we try the following encoding:
\begin{equation} \label{eq:3codewords}
\ket{\bar{0}} := \frac{1}{\sqrt{2}} (\ket{000} + \ket{111}) \qquad \ket{\bar{1}} := \frac{1}{\sqrt{2}} (\ket{000} - \ket{111})
\end{equation}
The states $\ket{\bar{0}}$ and $\ket{\bar{1}}$ are read as ``logical zero'' and ``logical one,'' and are often also called ``codewords.''

A useful feature of this encoding is that we can detect and correct a single erroneous bit flip, meaning that we can deduce whether $X_1$, $X_2$, or $X_3$ was applied to one of the physical qubits and then undo the damage.
For example, suppose that $X_1$ gets applied erroneously to one of the codewords:
\begin{equation}
X_1 \ket{\bar{0}} = \ket{100} + \ket{011} \qquad X_1 \ket{\bar{1}} = \ket{100} - \ket{011}
\end{equation}
(Here and henceforth, we will omit the factors of $1/\sqrt{2}$ to avoid cluttering the math in the rest of this section.)
Notice that $\ket{\bar{0}}$ and $\ket{\bar{1}}$ are eigenstates of the operators $Z_1 Z_2$ and $Z_2 Z_3$ with eigenvalue $+1$.
However, after applying $X_1$, we see that
\begin{equation}
\begin{aligned}
Z_1 Z_2 ( \ket{100} \pm \ket{011} ) &= - ( \ket{100} \pm \ket{011} ) \\
Z_2 Z_3 ( \ket{100} \pm \ket{011} ) &= + ( \ket{100} \pm \ket{011} )
\end{aligned}
\end{equation}
Similarly, if we measure $Z_1 Z_2$ and $Z_1 Z_3$ after applying $X_2$ or $X_3$, we can build up the following table:
\begin{center}
\begin{tabular}{ c | c c c }
~ & \multicolumn{3}{c}{error} \\
measurement & $X_1$ & $X_2$ & $X_3 $  \\
\hline
$Z_1 Z_2$ & $-1$ & $-1$ & $+1$ \\
$Z_2 Z_3$ & $+1$ & $-1$ & $-1$
\end{tabular}
\end{center}
Therefore, we can use the results of measuring $Z_1 Z_2$ and $Z_2 Z_3$ to deduce whether $X_1$, $X_2$, $X_3$, or no error occurred ($+1$ is obtained in both measurements).
Then, since $X_i^2 = I$, all we have to do is apply the right $X$ operator again to correct the error.

The codewords are also eigenstates of $Z_1 Z_3$ with eigenvalue $+1$, and they become eigenstates with eigenvalue $-1$ after a single bit flip error occurs.
This is not independent information, however, since $Z_1 Z_3 = (Z_1 Z_2)(Z_2 Z_3)$.
More generally, $\ket{\bar{0}}$ and $\ket{\bar{1}}$ are the $+1$ eigenstates of the \emph{group} generated by $Z_1 Z_2$ and $Z_2 Z_3$.
We call this group the \emph{stabilizer group}, $S$. $Z_1 Z_2$ and $Z_2 Z_3$ are called \emph{stabilizer generators}, and we write $S = \langle Z_1 Z_2, Z_2 Z_3 \rangle$.
This formalism generalizes in a powerful way, resulting in a class of QECCs that are called \emph{stabilizer codes}.
For an introduction to these codes, see \cite{gottesman}.

While we can correct a single bit flip, a single phase flip ($Z_1$, $Z_2$, or $Z_3$) on the other hand is bad news.
From \Eq{eq:3codewords}, we see that
\begin{equation}
Z_i \ket{\bar{0}} = \ket{\bar{1}} \qquad Z_i \ket{\bar{1}} = \ket{\bar{0}}.
\end{equation}
In other words, each $Z_i$ is a representation of a \emph{logical} $\bar{X}$ operator, and so a single erroneous phase flip results in a change of the logical state of the encoded qubit.
(Analogously, the logical $\bar{Z}$ operator is given by $\bar{Z} = X_1 X_2 X_3$.)
Unfortunately, this QECC is not very robust.

\subsubsection{The 9-qubit Shor code}

The 3-qubit code protected against a bit flip error, so perhaps more copies of this code can protect against a phase flip as well.
This is the gist of the 9-qubit Shor code \cite{PhysRevA.52.R2493,Devitt_2013}.
Suppose we have 9 physical qubits and that we encode our logical qubit as follows:
\begin{equation}\label{eq:9codewords}
\ket{\bar{0}} := (\ket{000} + \ket{111})^{\otimes 3} \qquad \ket{\bar{1}} := (\ket{000} - \ket{111})^{\otimes 3}
\end{equation}
Each codeword is made of three blocks, each of which is a copy of the corresponding 3-qubit codeword.
As such, we can detect a single bit flip in each block (for a total of up to 3 bit flips) by measuring the operators
\begin{equation} \label{eq:Zstab}
Z_1 Z_2, ~ Z_2 Z_3,~ Z_4 Z_5,~ Z_5 Z_6,~ Z_7 Z_8,~ Z_8 Z_9.
\end{equation}
This time, we can also detect a single phase flip (in total) by measuring the operators
\begin{equation} \label{eq:Xstab}
X_1 X_2 X_3 X_4 X_5 X_6, ~ X_4 X_5 X_6 X_7 X_8 X_9.
\end{equation}
Again notice that $\ket{\bar{0}}$ and $\ket{\bar{1}}$ are $+1$ eigenstates of these two operators.
Suppose, for example, that $Z_5$ gets applied erroneously.
Then for an arbitrary logical state, we have that
\begin{equation}
\begin{aligned}
X_1 X_2 X_3 X_4 X_5 X_6 (Z_5 (a \ket{\bar{0}} + b \ket{\bar{1}})) &= -Z_5 (a \ket{\bar{0}} + b \ket{\bar{1}}) \\
X_4 X_5 X_6 X_7 X_8 X_9 (Z_5 (a \ket{\bar{0}} + b \ket{\bar{1}})) &= -Z_5 (a \ket{\bar{0}} + b \ket{\bar{1}}).
\end{aligned}
\end{equation}
Proceeding similarly, we can build up a table as we did before:
\begin{center}
\begin{tabular}{ c | c c c }
~ & \multicolumn{3}{c}{error} \\
measurement & $Z_1$ or $Z_2$ or $Z_3$ & $Z_4$ or $Z_5$ or $Z_6$ & $Z_7$ or $Z_8$ or $Z_9$  \\
\hline
$X_1 X_2 X_3 X_4 X_5 X_6$ & $-1$ & $-1$ & $+1$ \\
$X_4 X_5 X_6 X_7 X_8 X_9$ & $+1$ & $-1$ & $-1$
\end{tabular}
\end{center}
Therefore, by measuring these two strings of $X$ operators, we can figure out in which block the phase flip occurred.
It does not matter that we are ignorant of which particular qubit experienced the $Z$ error, since applying $Z$ to any qubit in the right block will flip the phase back.

In the language of stabilizer codes, the stabilizer group is generated by the operators \eqref{eq:Zstab} and \eqref{eq:Xstab}.
We also see that representatives of the logical $\bar{Z}$ and $\bar{X}$ operators are
\begin{equation}
\bar{Z} = X_1 X_2 X_3 \qquad \bar{X} = Z_1 Z_4 Z_7.
\end{equation}
Multiplying these representations by elements of the stabilizer group produces different equivalent representations of the logical $\bar{Z}$ and $\bar{X}$ operators.

Already in these two examples we see the features of QECCs that we highlighted before.
These two codes encode logical information nonlocally across 3 and 9 qubits, respectively, and they cannot correct arbitrary errors.
Error diagnosis is always carried out by performing collective measurements that access several physical qubits at once.
It is imperative that we never make any local measurements so that we do not disturb the computational state.
Such measurements are typically made using extra \emph{ancillary} qubits.
For example, Exercise~5 discusses how to collectively and non-destructively measure $Z_1 Z_2$ and $X_1 X_2 X_3 X_4 X_5 X_6$.

\subsection{Quantum error correction as a quantum channel}

\begin{figure}[ht]
\centering
\includegraphics[width=\textwidth]{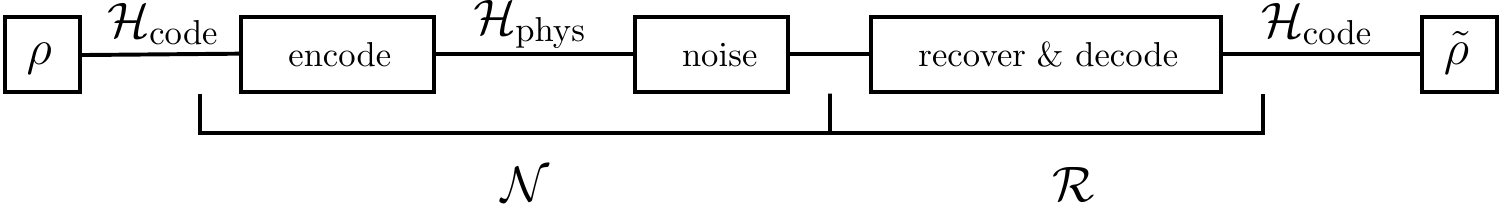}
\caption{Quantum error correction as a quantum channel.}
\label{fig:QECasChannel}
\end{figure}

Schematically, we can represent quantum error correction as a series of steps, as shown in \Fig{fig:QECasChannel}.
We start with some initial logical state $\rho \in \mathcal{S}(\Hil_\mrm{code})$ that we encode in $\mathcal{H}_\mrm{phys}$.
Noise then gets applied to the encoded state, which we then attempt to recover from and decode to get back to a logical state $\tilde \rho$.
For error correction to be successful, we must have $\tilde \rho \approx \rho$.

Each one of the steps in this process---encoding, noise, recovery, and decoding---is a quantum channel.
If we denote the encoding and noise steps by the channel $\mathcal{N}$ and the recovery and decoding steps by the channel $\mathcal{R}$, then the signature of successful error correction is
\begin{equation}
(\mathcal{R} \circ \mathcal{N})(\rho) \approx \rho.
\end{equation}
In other words, we want to \emph{reverse} the channel $\mathcal{N}$ as best as is possible.

\begin{figure}[ht]
\centering
\includegraphics[scale=1]{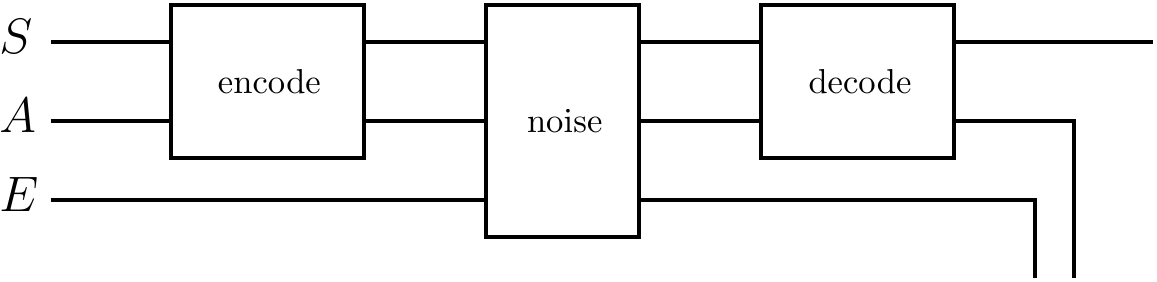}
\caption{Quantum error correction as a unitary process.}
\label{fig:QECunitary}
\end{figure}

What are the criteria that ensure that error correction will be successful?
In other words, given an encoding and noise channel $\mathcal{N}$, when does there exist a faithful recovery channel $\mathcal{R}$?
We can gain some heuristic intuition by considering the purified theory.
Recall from \Sec{sec:isometricdilation} that we can think of any quantum channel as being a unitary process in a larger Hilbert space.
If we call $\Hil_\mrm{code}$ the system, $S$, which we augment with ancillas, $A$, that are used in encoding and decoding, as well as an environment, $E$, that participates in the noisy interactions, then a unitary version of the error correction process is as shown in \Fig{fig:QECunitary}.
Heuristically, it must be that the final state of $E$ cannot depend on the initial state of $S$ in order for no information to be lost and for perfect recovery to be possible.

More precisely, the following theorem lays out when it is possible to exactly reverse a quantum channel \cite{ohya1993quantum}:
\begin{thm}[Petz and Ohya] \label{thm:Petz}
Let $\mathcal{N} : \mathcal{L}(\Hil_A) \rightarrow \mathcal{L}(\Hil_B)$ be a quantum channel and $Q \subseteq \mathcal{S}(\Hil_A)$.
$\mathcal{N}$ is exactly reversible on $Q$ if and only if
\begin{equation} \label{eq:recoveryCriterion} 
\rent{\rho}{\sigma} = \rent{\mathcal{N}(\rho)}{\mathcal{N}(\sigma)}
\end{equation}
for all $\rho, \sigma \in Q$.
Furthermore, for any $\sigma \in Q$ such that $\mrm{supp} \, \rho \subseteq \mrm{supp} \, \sigma$ for all $\rho \in Q$, a channel that undoes the action of $\mathcal{N}$ is
\begin{equation}
\mathcal{P}_{\mathcal{\sigma},\mathcal{N}} ~ : ~ \rho ~ \mapsto ~ \sigma^{1/2} \mathcal{N}^\dagger \left[ \mathcal{N}(\sigma)^{-1/2} \, \rho \, \mathcal{N}(\sigma)^{-1/2}  \right] \sigma^{1/2}.
\end{equation}
\end{thm}

The channel $\mathcal{P}_{\mathcal{\sigma},\mathcal{N}}$ is called the \emph{Petz map}.
While the Petz map is somewhat complicated, the criterion for exact recovery, $\rent{\rho}{\sigma} = \rent{\mathcal{N}(\rho)}{\mathcal{N}(\sigma)}$, has a clear meaning in light of Thm.~\ref{thm:Dmono}:
A channel $\mathcal{N}$ is only reversible for a collection of states when $\mathcal{N}$ does not reduce their distinguishability.

A complete proof of Petz and Ohya's theorem is well beyond what we can succinctly accomplish here.
If you are interested in seeing the proof, Chapter 12 of \Ref{Wilde:2011npi} is largely devoted to this.
It's almost trivial that $(\mathcal{P}_{\mathcal{\sigma},\mathcal{N}} \circ \mathcal{N})(\sigma) = \sigma$ (the only missing step is showing that $\mathcal{N}^\dagger(I) = I$).
The harder part is showing that $(\mathcal{P}_{\mathcal{\sigma},\mathcal{N}} \circ \mathcal{N})(\rho) = \rho$ for other $\rho \in Q$.
Instead, let's check that the Petz map works for a specific simple example.\footnote{This example is inspired by an example that appears in \Ref{Penington:2019kki} to illustrate a novel kind of entanglement wedge reconstruction in AdS/CFT.}

\begin{eg} \label{eg:Petz}
Let $\dim \Hil_\mrm{code} = d_\mrm{code}$ and suppose that we use an isometry, $V$, to embed $\Hil_\mrm{code}$ into a larger Hilbert space with the tensor product structure $\Hil_\mrm{phys} = \Hil_A \otimes \Hil_{\bar{A}}$.
Explicitly,
\begin{equation}
V : \Hil_\mrm{code} \rightarrow \Hil_A \otimes \Hil_{\bar{A}}, \qquad V^\dagger V = I_\mrm{code}, \qquad \text{and} \qquad VV^\dagger = \Pi_\mrm{code},
\end{equation}
where $\Pi_\mrm{code}$ is the projector onto $V(\Hil_\mrm{code}) \subset\Hil_A \otimes \Hil_{\bar{A}}$.
Define a channel
\begin{equation}
\begin{aligned}
\mathcal{N} ~ : ~ \mathcal{S}(\Hil_\mrm{code}) ~ &\rightarrow ~ \mathcal{S}(\Hil_A) \\
 \rho ~ &\mapsto ~ \Tr_{\bar{A}}(V \rho V^\dagger),
\end{aligned}
\end{equation}
which embeds a code state $\rho$ into $\Hil_A \otimes \Hil_{\bar{A}}$ and then ``erases'' $\Hil_{\bar{A}}$.
Let's also fix a full-rank fiducial state $\sigma \in \mathcal{S}(\Hil_\mrm{code})$, i.e., letting $\{\ket{a}_\mrm{code}\}_{a=1}^{d_\mrm{code}}$ be a basis for $\Hil_\mrm{code}$, pick a state $\sigma = \sum_{a=1}^{d_\mrm{code}} \sigma_a \ketbra{a}{a}$ with each $\sigma_a \neq 0$.

For exact recovery to be possible on all of $\Hil_\mrm{code}$, information cannot leak into $\bar{A}$ and become lost when we trace this factor out.
Therefore, in a setting where exact recovery is possible, we must have
\begin{equation}
\Hil_A \cong \Hil_1 \otimes \Hil_2 \oplus \Hil_3,
\end{equation}
where $\dim \Hil_1 \equiv d_1 = d_\mrm{code}$, $\dim \Hil_2 \equiv d_2 \geq 1$, and $\dim \Hil_3 \equiv d_3 \geq 0$.
(The space $\Hil_3$ plays no other role than to make sure that the dimensions $d_1 d_2 + d_3$ add up to $d_A$.)
In this case, we can choose a basis of $\Hil_A \otimes \Hil_{\bar A}$ such that
\begin{equation}
V \ket{a}_\mrm{code} = \ket{a}_1 \otimes \ket{\chi}_{2\bar{A}},
\end{equation}
where $\ket{\chi}_{2\bar{A}} \in \Hil_2 \otimes \Hil_{\bar{A}}$ is the same fixed state for every $1 \leq a \leq d_\mrm{code}$.

Now, let's piece together the action of the Petz map.
We first evaluate $\mathcal{N}(\sigma)$:
\begin{align*}
\mathcal{N}(\sigma) &= \Tr_{\bar{A}} \left[ \sum_{a=1}^{d_\mrm{code}} \sigma_a \ketbra{a}{a}_1 \otimes \ketbra{\chi}{\chi}_{2\bar{A}} \right] \\
&= \left( \sum_{a=1}^{d_\mrm{code}} \sigma_a \ketbra{a}{a}_1 \right) \otimes \Tr_{\bar{A}} \ketbra{\chi}{\chi}_{2\bar{A}} \\
&\equiv \sigma_1 \otimes \chi_2
\end{align*}
Therefore, it follows that $\mathcal{N}(\sigma)^{-1/2} = \sigma_1^{-1/2} \otimes \chi_2^{-1/2}$.
Similarly, for an arbitrary state $\rho = \sum_{b,c=1}^{d_\mrm{code}} \rho_{bc} \ketbra{b}{c}_\mrm{code}$, one finds that $\mathcal{N}(\rho) = \rho_1 \otimes \chi_2$.
We therefore arrive at
\begin{equation}
\mathcal{N}(\sigma)^{-1/2} \mathcal{N}(\rho) \mathcal{N}(\sigma)^{-1/2} = (\sigma^{-1/2} \rho \sigma^{-1/2})_1 \otimes I_2.
\end{equation}

The next step is to deduce the action of $\mathcal{N}^\dagger$.
Let $\langle M, N \rangle \equiv \Tr(M^\dagger N)$ denote the operator inner product.
Letting $\tau, \omega \in \mathcal{S}(\Hil_\mrm{code})$, from the definition of the adjoint, we have:
\begin{align*}
\langle \mathcal{N}(\tau), \omega_1 \otimes I_2 \rangle_A &= \Tr_A[(\tau_1 \otimes \chi_2)(\omega_1 \otimes I_2)] \\
&= \Tr_1(\tau \omega) \Tr_2(\chi) \\
&= \Tr_\mrm{code} \left[ \tau  \left([\Tr \chi] \omega \right)\right] \\
&= \langle \tau, (\Tr \chi) \omega \rangle_\mrm{code}
\end{align*}
Therefore, $\mathcal{N}^\dagger(\omega_1 \otimes I_2) = (\Tr \chi) \omega_\mrm{code}$.

Putting it all together, we therefore have that
\begin{equation}
\begin{aligned}
\mathcal{P}_{\sigma, \mathcal{N}}[ \mathcal{N}(\rho) ] &= \sigma^{1/2} (\sigma^{-1/2} \rho \sigma^{-1/2})\sigma^{1/2} \cdot \Tr \chi \\
&= \rho \cdot \Tr \chi
\end{aligned}
\end{equation}
But, notice that
\begin{equation}
\Tr \chi = \Tr_2( \Tr_{\bar{A}} [\ketbra{\chi}{\chi}_{2\bar{A}}] ) = \braket{\chi}{\chi}_{2\bar{A}} = 1,
\end{equation}
and so we indeed find that $\mathcal{P}_{\sigma, \mathcal{N}}[ \mathcal{N}(\rho) ] = \rho$.

Since the Petz map perfectly recovers all states in $\Hil_\mrm{code}$, Petz and Ohya's theorem guarantees that the recoverability condition \eqref{eq:recoveryCriterion} holds.
Nevertheless, this can also be verified by direct computation:
\begin{equation}
\rent{\mathcal{N}(\rho)}{\mathcal{N}(\sigma)} = \rent{\rho_1 \otimes \chi_2}{\sigma_1 \otimes \chi_2} = \rent{\rho}{\sigma}
\end{equation}
The last equality is derived in Exercise~6 in \Sec{sec:exercises}.
\eop
\end{eg}

The Petz map is a remarkable constructive result; however, a limitation of Thm.~\ref{thm:Petz} is that it only lays out criteria for when a channel can be perfectly reversed.
If the condition \eqref{eq:recoveryCriterion} does not hold or only approximately holds, then Thm.~\ref{thm:Petz} does not say if and how well the Petz map will work.
The following theorem of Junge, Renner, Sutter, Wilde, and Winter lays out precisely this refinement \cite{Junge:2015lmb}:
\begin{thm}[Universal Recovery] \label{thm:universal}
Let $\mathcal{N} : \mathcal{L}(\Hil_A) \rightarrow \mathcal{L}(\Hil_B)$ be a quantum channel.
For all $\rho, \sigma \in \mathcal{S}(\Hil_A)$ such that $\mrm{supp} \, \rho \subseteq \mrm{supp} \, \sigma$, the recovery channel
\begin{equation}
\mathcal{R}_{\sigma, \mathcal{N}}(\rho) = \int_\mathbb{R} dt ~ \beta_0(t) \sigma^{-it/2} \mathcal{P}_{\sigma,\mathcal{N}}\left[ \mathcal{N}(\sigma)^{it/2} \, \rho \, \mathcal{N}(\sigma)^{-it/2}  \right] \sigma^{it/2},
\end{equation}
where $\beta_0(t) = \tfrac{\pi}{2} (\cosh(\pi t) + 1)^{-1}$, satisfies
\begin{equation} \label{eq:universalBound}
\rent{\rho}{\sigma} - \rent{\mathcal{N}(\rho)}{\mathcal{N}(\sigma)} \geq -2 \log F(\rho, \mathcal{R}_{\sigma, \mathcal{N}} \circ \mathcal{N}[\rho]).
\end{equation}
\end{thm}
The function $F(\rho,\sigma) = \Vert \rho^{1/2} \sigma^{1/2} \Vert_1$ is known as \emph{fidelity}. 
It is another measure of the closeness of two states, taking values between 0 and 1 and saturating at $F(\rho,\rho) = 1$.
The theorem above therefore says that the faithfulness with which $\mathcal{R}_{\sigma, \mathcal{N}}$ succeeds in reversing the action of $\mathcal{N}$ is upper bounded by the exact recoverability condition \eqref{eq:recoveryCriterion}.
In other words, the less a channel degrades the distinguishability of states on which it acts, the better its action can be reversed for these states.
The map $\mathcal{R}_{\sigma, \mathcal{N}}$ is known as a \emph{universal recovery channel}.

\section{An application to holography}
\label{sec:holo}

The Anti de Sitter/Conformal Field Theory (AdS/CFT) correspondence is a remarkable duality between certain gravitational theories and certain quantum field theories without gravity \cite{Maldacena:1997re,Witten:1998qj}.
In that sense, AdS/CFT is a genuine theory of quantum gravity, and so even if the gravitational side of the duality differs somewhat from the space-time of our Universe as we know it, AdS/CFT remains a window into quantum gravity.

In these notes, we will take AdS/CFT to mean the following:
\begin{quote}
Among certain quantum field theories without gravity in $d$ space-time dimensions called conformal field theories (CFTs), certain CFTs are exactly equivalent to quantum theories of asymptotically Anti de Sitter (AdS) space-times in $d+1$ dimensions.
Moreover, in the right limit, certain CFT states are in exact correspondence with certain fixed asymptotically AdS space-times.
\end{quote}
I like to think of the definition above as ``AdS/CFT: the conjecture,'' which is to be distinguished from ``AdS/CFT: the theorem.''
The latter refers to the precise and rigorous correspondence between specific superconformal field theories and specific string theories in specific limits and in specific numbers of dimensions.
The former envisions a much broader scope of applicability, but has correspondingly less backing by formal calculations in string theory and conformal field theory.
Nevertheless, the broader applicability has made it possible to use tools and techniques from quantum information to study the correspondence, which has given us deep information-theoretic insights into AdS/CFT, and more generally (we think) quantum gravity itself.

We will not go into any precise details of AdS/CFT in these notes.
All we will do is illustrate the basic idea of the duality with a simple example and then point out the information-theoretic connection.
For a more comprehensive introduction to AdS/CFT, \Ref{Nastase:2007kj} is one place you could start.
Afterwards, we will see how the information-theoretic tools that we have developed can be used to relate operators on the gravitational side of the duality to corresponding operators in the dual field theory.

\subsection{How to bluff your way through AdS/CFT}

A CFT is a quantum field theory with a specific set of symmetries (namely, conformal symmetry) which make it so that there is no inherent absolute notion of scale in the theory.
We say that a CFT is \emph{holographic} when it has a dual gravitational description in terms of asymptotically AdS space-times.
In the simplest case, the ground state of a holographic CFT in $d$ space-time dimensions is dual to pure $(d+1)$-dimensional AdS space-time.

\begin{figure}[ht]
\centering
\includegraphics[scale=1]{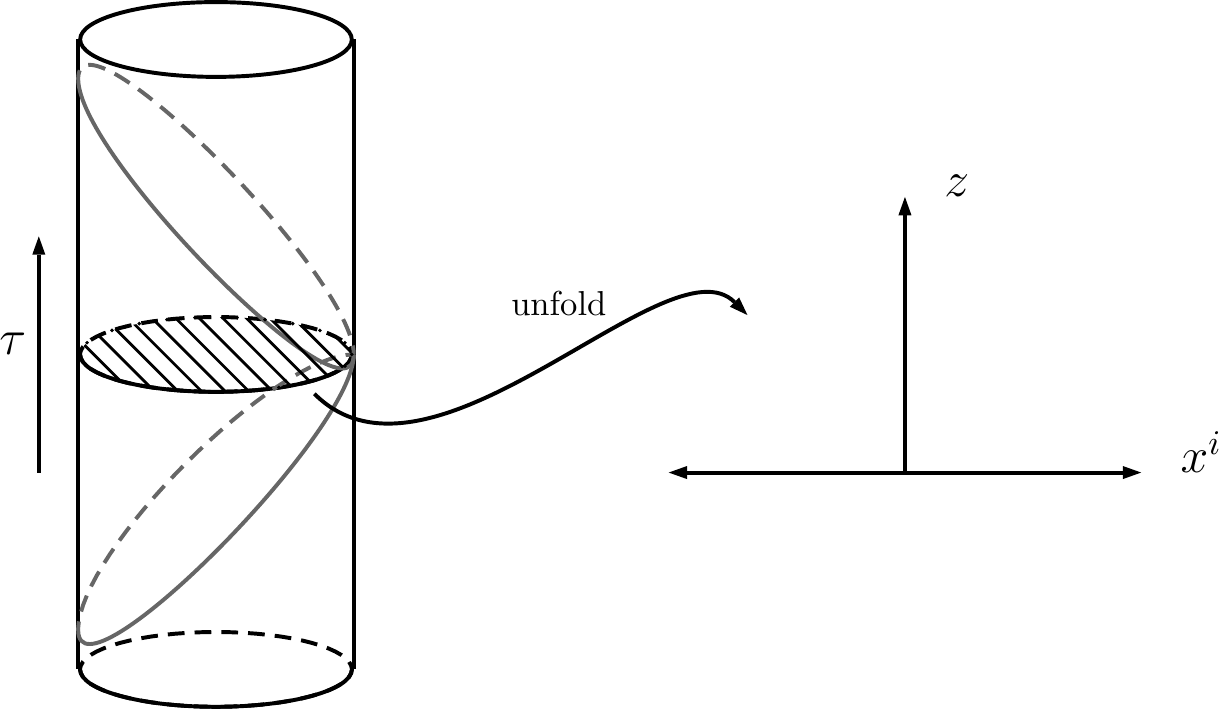}
\caption{Anti de Sitter space-time. Slices of the global cylinder are hyperbolic spaces. The Poincar{\'e} patch is the space-time in between the two tilted circles. Unfolding the $\tau = 0$ slice gives us the a slice of the Poincar{\'e} patch in planar coordinates, where proper distances increase as one approaches the boundary at $z=0$.}
\label{fig:AdS}
\end{figure}

AdS is a maximally symmetric space-time with constant negative curvature.
In an appropriate set of coordinates, one way to visualize $\text{AdS}_{d+1}$ is as a cylinder, as shown in \Fig{fig:AdS}.
Time $\tau$ runs up along the cylinder, and slices of the cylinder are $d$-dimensional hyperbolic spaces.
The \emph{Poincar{\'e} patch} is only a portion of $\text{AdS}_{d+1}$, but it is covered by a simple set of coordinates that makes the geometry easy to understand:
\begin{equation} \label{eq:poincareAdS}
ds^2 = \frac{L^2}{z^2} \left( -dt^2 + dz^2 + dx_i dx^i \right)
\end{equation}
$L$ is called the AdS length, and it is related to the cosmological constant by
\begin{equation}
\Lambda = - \frac{d(d-1)}{2 L^2}.
\end{equation}
If we take the $\tau = 0$ slice of the cylinder in \Fig{fig:AdS} (which coincides with $t=0$) and imagine making an incision at a point on its boundary, then we can unfold the slice into an upper half-plane as shown in the right side of \Fig{fig:AdS}.
The coordinate $z$ starts at $z = 0$ at the boundary and increases as we move into the AdS bulk, and the $x^i$ coordinates are parallel to the boundary.
In this plane, the interpretation of the line element \eqref{eq:poincareAdS} is clear:
Setting $dt = dz = 0$, we see that a small fixed coordinate displacement $dx^i$ has larger proper length the closer we are to the AdS boundary.
Notice that the AdS boundary has $d$ space-time dimensions.
As such, it is often very convenient to think of the dual CFT as living on the AdS boundary.

In general, the connection that quantum information has to AdS/CFT is that \emph{information-theoretic quantities in the boundary CFT correspond to geometric quantities in the AdS bulk}.
The most basic example of this is a formula that relates the entanglement entropy of a reduced state in the boundary CFT to the area of an extremal surface in the AdS bulk.

\begin{figure}[ht]
\centering
\includegraphics[scale=1]{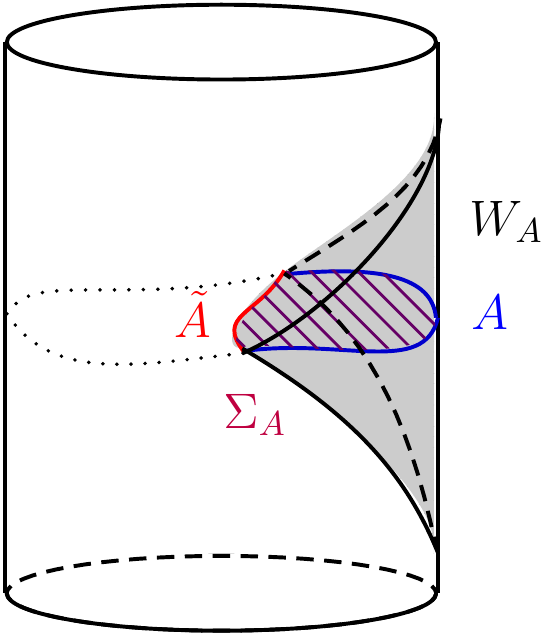}
\caption{Various geometric objects: a boundary subregion, $A$ (blue); the HRT surface, $\tilde A$ (red); the entanglement wedge, $W_A$ (shaded grey); and a complete spacelike slice through the entanglement wedge, $\Sigma_A$, such that $\partial \Sigma_A = A \cup \tilde{A}$ (hatched purple).}
\label{fig:HRT}
\end{figure}

Let $A$ be a (spacelike) subregion of a holographic boundary CFT state that is dual to a fixed asymptotically AdS space-time, and let $\rho_A$ be the reduced CFT state on this subregion.
Then, its entropy is given by
\begin{equation} \label{eq:HRT}
S(\rho_A) = \min \underset{A \sim \tilde{A}}{\mrm{ext}} \frac{\mrm{area}(\tilde{A})}{4 G_N} + O(G_N^0).
\end{equation}
The formula above says that we look for co-dimension 2 spacelike surfaces $\tilde A$ in the bulk that are \emph{extremal}, meaning that their area is locally stationary under null variations (equivalently, the expansions of orthogonal null congruences anchored to the boundary of $\tilde A$ vanish).
Furthermore, $A \sim \tilde A$ denotes that $\tilde A$ must be \emph{homologous} to $A$, meaning that it can be smoothly deformed into $A$.
Then, if there are many such surfaces $\tilde A$, we take the one with the smallest area, and $1/4$ of this area in Planck units gives $S(\rho_A)$.
\Fig{fig:HRT} illustrates this geometry.

\Eq{eq:HRT} is known as the \emph{Hubeny-Rangamani-Takayanagi} (HRT) formula, and the smallest-area extremal surface is called the \emph{HRT surface of $A$} \cite{Hubeny:2007xt}.
As a historical note, this is a refinement of the original entropy formula due to Ryu and Takayanagi (RT), which is applicable to the case where the dual space-time is static \cite{Ryu:2006ef}.
In this case, one only needs to look for minimal surfaces in a spacelike slice of the space-time, and so the RT formula more simply reads
\begin{equation}
S(\rho_A) = \min_{A \sim \tilde A} \frac{\mrm{area}(\tilde A)}{4 G_N} + O(G_N^0).
\end{equation}

\subsection{Bulk reconstruction}

If AdS/CFT is to be a true duality, then \emph{any} quantity in the bulk AdS must be encoded somehow in the boundary CFT.
Naturally, then, we might ask: What do bulk operators look like in the boundary CFT, or equivalently, how do we reconstruct bulk operators using boundary CFT operators?
This is the subject of \emph{bulk reconstruction}.

\subsubsection{The extrapolate dictionary}

One of the earliest answers to this question was given by Hamilton, Kabat, Lifshytz, and Lowe (HKLL) for a free scalar field $\phi$ of mass $m$ in $\text{AdS}_{d+1}$ \cite{Hamilton:2006az}.
HKLL is based on the ``extrapolate dictionary,''
\begin{equation} \label{eq:HKLL}
\lim_{r \rightarrow \infty} r^\Delta \phi(r,t,x^i) = \Oh(t,x^i),
\end{equation}
where $r \propto z^{-1}$ so that $r \rightarrow \infty$ is the AdS boundary.
The extrapolate dictionary basically says that $\phi$ is in correspondence with an operator $\Oh$ in the boundary CFT (a primary operator with scaling dimension $\Delta$ that is related to $m$, $L$, and $d$) if you push $\phi$ to the boundary while weighting it with a factor $r^\Delta$.
You would be right to think that \Eq{eq:HKLL} is a bit incongruous, since it looks like we are equating a bulk AdS operator on the left side with a boundary CFT operator on the right side.
More correctly, the basic strategy is to look for a CFT operator $\tilde \phi$ that satisfies \eqref{eq:HKLL} (with $\phi \rightarrow \tilde \phi$, of course) as well as an equation of motion
\begin{equation} \label{eq:EOM}
(\Box - m^2) \tilde \phi = 0,
\end{equation}
where $\Box$ is the scalar d'Alembertian coming from the bulk theory for $\phi$.
In other words, $\tilde \phi$ is a CFT operator that depends on the boundary coordinates $(t,x^i)$, but that also has an additional parameter $r$ so that it altogether satisfies \Eqs{eq:HKLL}{eq:EOM}.

HKLL showed that such an operator may be expressed as
\begin{equation}
\tilde \phi(x) = \int dX K(x,X) \Oh(X),
\end{equation}
where $x \equiv (r,t,x^i)$ denotes a bulk point and $X \equiv (t,x^i)$ denotes a boundary point.
The function $K(x,X)$ is known as the \emph{smearing function}, and it ends up being expressed in terms of the mode functions of $\phi$.
The boundary operator $\tilde \phi$ succeeds in reconstructing the bulk operator $\phi$ in the sense that boundary expectation values of $\tilde \phi$ reproduce the bulk expectation values of $\phi$:
\begin{equation}
\langle \phi(x_1) \phi(x_2) \cdots \rangle_\mrm{AdS} = \langle \tilde \phi(x_1) \tilde \phi(x_2) \cdots \rangle_\mrm{CFT}
\end{equation}

\begin{figure}[ht]
\centering
\includegraphics[scale=1]{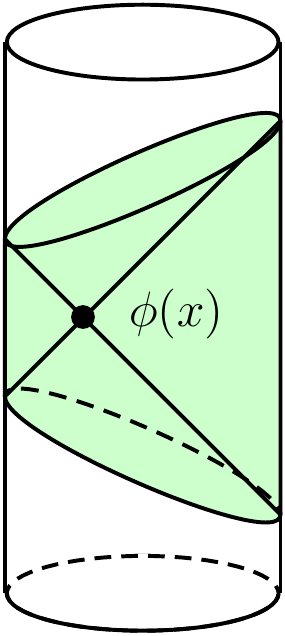}
\caption{In the HKLL reconstruction of a bulk operator $\phi(x)$, only boundary points that are spacelike-separated from $x$ contribute to the reconstruction, shown shaded.}
\label{fig:HKLL}
\end{figure}

The smearing function $K(x,X)$ has the property that it is only non-zero on boundary points that are spacelike-separated from $x$, as shown in \Fig{fig:HKLL}.
If we pick a single Cauchy slice $\Gamma$ of the boundary within the support of $K$, then it turns out that it's possible to propagate $K(x,X)$ backwards and forwards towards this slice to obtain a new smearing function that only has support on $\Gamma$.
In some sense, this results in a more ``efficient'' boundary representation of $\phi$.
In the next subsection, we will see that entanglement wedge reconstruction leads to even more efficient representations.

\subsubsection{Entanglement wedge reconstruction and error correction}

Let's begin by defining the entanglement wedge.
\begin{defn}
Given a boundary subregion $A$ with a HRT surface $\tilde A$, the \emph{entanglement wedge} of $A$, denoted $W_A$, is the bulk domain of dependence of any spacelike surface $\Sigma_A$ such that $\partial \Sigma_A = A \cup \tilde{A}$.
\end{defn}
\Note The domain of dependence of $\Sigma_A$ is the collection of points $p$ such that any causal curve through $p$ intersects $\Sigma_A$, and $\partial \Sigma_A$ denotes the boundary of $\Sigma_A$.
\medskip

\noindent An example of an entanglement wedge is illustrated in \Fig{fig:HRT} above.

The entanglement wedge is important for bulk reconstruction because if a bulk operator has support on $W_A$, then it can be represented by a CFT operator that has support only on $A$.
This principle is known as \emph{entanglement wedge reconstruction}.
Moreover, it establishes a notion of equivalence between specific bulk and boundary regions that we call \emph{subregion-subregion duality}.
In the sense of bulk reconstruction at least, a boundary subregion $A$ is dual to its entanglement wedge $W_A$ in the bulk.
This characterization of entanglement wedge reconstruction is fairly imprecise, but we will look at a much more careful and precise version in the next subsection.

\begin{figure}[ht]
\centering
\includegraphics[scale=1]{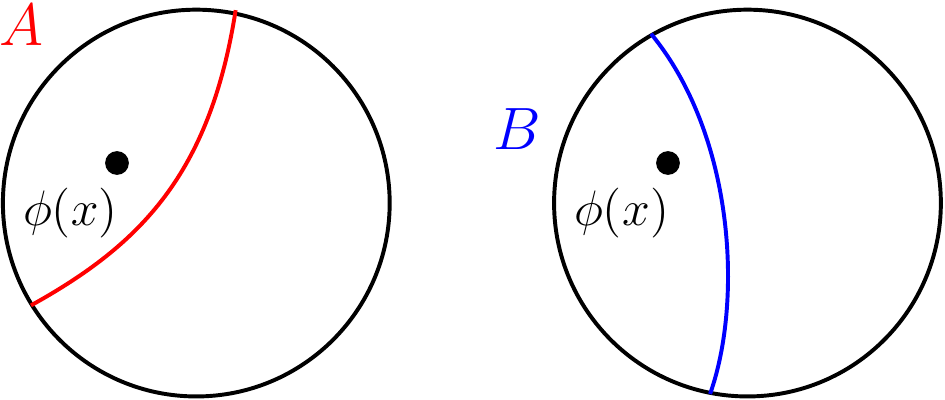}
\caption{A single operator $\phi(x)$ is in the entanglement wedge of infinitely many boundary subregions.}
\label{fig:holoQECC}
\end{figure}

We should note that entanglement wedge reconstruction raises a question of interpretation that we will also address precisely.
For any given bulk operator whose support is not the entire bulk, then there is no unique boundary subregion whose entanglement wedge contains that operator, as shown in \Fig{fig:holoQECC}.
It would then seem that it's possible to represent the same bulk operator with different CFT operators on different boundary subregions that need not have any overlap.
In what sense are these different CFT operators the ``same''?

The answer that we will substantiate in the next section is that we can think of the encoding of bulk operators in the CFT boundary as a quantum error correcting code that protects against deletion of portions of the boundary.
The bulk is encoded nonlocally and redundantly in the boundary such that we can recover a given bulk operator provided we hold enough of the boundary.
Moreover, once a bulk operator has been encoded in the boundary via AdS/CFT, we can think of its reconstruction as a CFT operator on a subregion $A$ as a recovery map on $A$ after having discarded the complementary region $\bar{A}$, just as in Ex.~\ref{eg:Petz}.

\subsubsection{Bulk reconstruction as a universal recovery channel}

Currently, the most precise characterization of entanglement wedge reconstruction is the following one, due to Cotler, Hayden, Penington, Salton, Swingle, and Walter \cite{Cotler:2017erl}.

\begin{figure}[ht]
\centering
\includegraphics[scale=1]{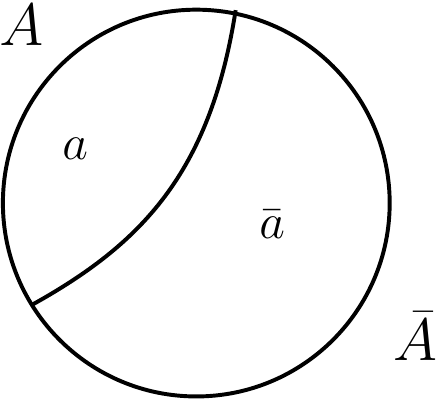}
\caption{Operators in $a$ can be reconstructed on $A$.}
\label{fig:bulkReconstruction}
\end{figure}

Given a holographic CFT and a boundary subregion $A$, write $\Hil_\mrm{CFT} = \Hil_A \otimes \Hil_{\bar{A}}$, and correspondingly factorize $\Hil_\mrm{bulk} = \Hil_a \otimes \Hil_{\bar a}$, where $a$ is a Cauchy surface for $W_A$ and $a \cup \bar{a}$ is a Cauchy surface for the entire bulk.\footnote{Following Cotler \emph{et al.}, we assume that the Hilbert spaces factorize for convenience. This is not a given for a holographic CFT; however, all of their arguments can be made more carefully at the level of operator algebras without assuming Hilbert space factorization.}
This is illustrated in \Fig{fig:bulkReconstruction}.
Let $\Hil_\mrm{code} \subset \Hil_\mrm{bulk}$ be generated by a finite collection of states that have the same dual bulk geometry in a neighbourhood of $a$ up to corrections that are $O(\sqrt{G_N})$ in size (so that the bulk factorization makes sense for all code states, among other reasons).
We assume that the AdS/CFT correspondence supplies us with an isometry $J : \Hil_\mrm{code} \hookrightarrow \Hil_\mrm{CFT}$ that embeds $\Hil_\mrm{code}$ into $\Hil_\mrm{CFT}$.
Then, given a bulk operator $\phi_a \in \mathcal{L}(\Hil_a)$, our goal is to find a CFT operator $\Oh_A \in \mathcal{L}(\Hil_A)$ such that
\begin{equation} \label{eq:CHPSSW}
\left| \langle \Oh_A \rangle_{J\rho J^\dagger} - \langle \phi_a \rangle_\rho  \right| \leq \delta \Vert \phi_a \Vert
\end{equation}
for all $\rho \in \mathcal{S}(\Hil_\mrm{code})$.
$\langle \Oh_A \rangle_{J\rho J^\dagger}$ denotes a CFT expectation value with respect to the state $J\rho J^\dagger$, $\langle \phi_a \rangle_\rho$ denotes a bulk expectation value with respect to $\rho$, $\delta$ is a small fixed constant, and $\Vert \phi_a \Vert \equiv \max_{\Vert v \Vert = 1} \Vert \phi_a v \Vert$ is the operator norm of $\phi_a$.
We will call success in this task ``entanglement wedge reconstruction.''

There is an important additional result from AdS/CFT due to Jafferis, Lewkowycz, Maldacena, and Suh (JLMS) \cite{Jafferis:2015del} that we will need in order to bound the left side of \eqref{eq:CHPSSW}.
In terms of the language that we are using here, the result reads
\begin{equation}\label{eq:JLMS}
|\rent{\rho_A}{\sigma_A} - \rent{\rho_a}{\sigma_a}| \leq O(\sqrt{G_N}) \qquad \forall ~ \rho, \sigma \in \mathcal{S}(\Hil_\mrm{code}),
\end{equation}
where $\rho_a = \Tr_{\bar{a}} \rho$ and $\rho_A = \Tr_{\bar{A}}[J \rho J^\dagger]$ (and similarly for $\sigma$).
In particular, this result is what fixes the region that we can reconstruct, $a$, to be the entanglement wedge of $A$.

The first strategy that comes to mind is to try mimicking what we did in Ex.~\ref{eg:Petz}.
If we define a channel $\tilde{\mathcal{N}}(\rho) = \Tr_{\bar{A}}[J \rho J^\dagger]$, then we can write down a universal recovery channel for it.
Then, applying the bound from Thm.~\ref{thm:universal}, perhaps we can use the JLMS bound on relative entropy to arrive at the desired inequality \eqref{eq:CHPSSW}?

A problem with this simple strategy is that $\Tr_{\bar{A}}[J \rho J^\dagger]$ in principle depends on $\bar{a}$ as well.
If we are really after reconstruction in the \emph{entanglement wedge}, then all of our results had better only depend on states defined on $a$ and $A$.
While we therefore cannot immediately declare victory, it turns out that only minor modifications are needed to get an approach that works.

The proof proceeds in 3 steps.
First, one defines a channel
\begin{equation}
\mathcal{N}(\rho_a) = \Tr_{\bar A} \left[ J \, (\rho_a \otimes \bar{\sigma}_{\bar{a}}) \, J^\dagger \right],
\end{equation}
where $\bar{\sigma}_{\bar{a}} \in \mathcal{S}(\Hil_{\bar{a}})$ is some fixed full-rank state that we choose.
Choosing another full-rank state $\sigma_a \in \mathcal{S}(\Hil_a)$, Thm.~\ref{thm:universal} supplies us with a universal recovery map $\mathcal{R}_{\sigma_{a}, \mathcal{N}}$ such that \Eqs{eq:universalBound}{eq:JLMS} give us
\begin{equation} \label{eq:bound1}
-2 \log F(\rho_a, \mathcal{R}_{\sigma_a, \mathcal{N}} \circ \mathcal{N} [\rho_a]) \leq |\rent{\rho_A}{\sigma_A} - \rent{\rho_a}{\sigma_a}| \leq \epsilon,
\end{equation}
for all $\rho \in \mathcal{S}(\Hil_\mrm{code})$ of the form $\rho_a \otimes \bar{\sigma}_{\bar{a}}$, and where $\epsilon$ is a fixed constant of size $O(\sqrt{G_N})$.
Second, one shows that this channel $\mathcal{R}_{\sigma_{a}, \mathcal{N}}$ still succeeds in reversing $\mathcal{N}$ for arbitrary $\rho \in \mathcal{S}(\Hil_\mrm{code})$ by using the bound \eqref{eq:bound1} to show that
\begin{equation}
\Vert \, \rho_a - \mathcal{R}[\Tr_{\bar{A}}(J \rho J^\dagger)] \, \Vert_1 \leq \delta,
\end{equation}
where we have dropped the subscript on $\mathcal{R}$ for neatness and where $\delta$ is another parametrically small constant that depends on $\epsilon$.
(In other words, here we start with an arbitrary $\rho \in \mathcal{S}(\Hil_\mrm{code})$, trace out $\bar{a}$ to obtain $\rho_a$, and then try to recover $\mathcal{N}(\rho_a)$ with $\mathcal{R}$, which is \emph{a priori} only guaranteed to work well had $\rho$ taken the specific form $\rho = \rho_a \otimes \bar{\sigma}_{\bar{a}}$.)
Third, one defines the operator
\begin{equation}
\Oh_A = \mathcal{R}^\dagger[\phi_a]
\end{equation}
and shows that it satisfies \Eq{eq:CHPSSW}.

The technical steps of the proof are not too difficult to follow either, provided that you are willing to refer to a couple of other sources for the proofs of some inequalities.
The calculation given here is essentially verbatim the calculation from Cotler \emph{et al.} \cite{Cotler:2017erl}, although I have explained a handful of inequalities to make these notes self-contained.

First, with the bound \eqref{eq:bound1} in hand, one uses the Fuchs-Van de Graaf inequality \cite{761271} to show that
\begin{equation} \label{eq:step1a}
\Vert \rho_a - \mathcal{R}(\mathcal{N}[\rho_a]) \Vert_1 \leq 2 \sqrt{\epsilon} \equiv \delta_1 .
\end{equation}
for all $\rho_a \in \mathcal{S}(\Hil_a)$.

To attack the second step, let $\rho \in \mathcal{S}(\Hil_\mrm{code})$, $\rho_a = \Tr_{\bar A} \rho$, and observe that
\begin{align}
\nonumber \Vert \mathcal{N}(\rho_a) - (J \rho J^\dagger)_A \Vert_1^2 &= \Vert (J \, \rho_a \otimes \bar{\sigma}_a \, J^\dagger)_A - (J \rho J^\dagger)_A \Vert_1^2 \\
&\leq (2 \log 2) \rent{(J \, \rho_a \otimes \bar{\sigma}_a \, J^\dagger)_A}{(J \rho J^\dagger)_A} . \label{eq:step2a}
\end{align}
Following Cotler \emph{et. al}, we use a bracket with a subscript to denote a partial trace, i.e., $(\Oh)_A \equiv \Tr_{\bar A}(\Oh)$.
To go to the second line, we used Pinsker's inequality \eqref{eq:Pinsker}.
Next, we apply JLMS to a ``trivial'' case to obtain
\begin{equation} \label{eq:step2b}
| \rent{\rho_a}{\rho_a} - \rent{(J \, \rho_a \otimes \bar{\sigma}_a \, J^\dagger)_A}{(J \rho J^\dagger)_A} | \leq \epsilon .
\end{equation}
$\rent{\rho_a}{\rho_a} = 0$ of course, and so letting $(2 \log 2) \epsilon \equiv \delta_2^2$, we combine \Eqs{eq:step2a}{eq:step2b} to obtain
\begin{equation} \label{eq:step2c}
\Vert \mathcal{N}(\rho_a) - (J \rho J^\dagger)_A \Vert_1 \leq \delta_2 .
\end{equation}
Finally, we have the following calculation:
\begin{align*}
\Vert \, \rho_a - \mathcal{R}[(J \rho J^\dagger)_A] \, \Vert_1 & \leq \Vert \, \rho_a - \mathcal{R}(\mathcal{N}[\rho_a]) \, \Vert_1 + \Vert \, \mathcal{R}(\mathcal{N}[\rho_a]) - \mathcal{R}[(J\rho J^\dagger)_A] \, \Vert_1 \\
&\leq  \Vert \, \rho_a - \mathcal{R}(\mathcal{N}[\rho_a]) \, \Vert_1 + \Vert \, \mathcal{N}(\rho_a) - (J\rho J^\dagger)_A \, \Vert_1 \\
&\leq \delta_1 + \delta_2 \equiv \delta
\end{align*}
In the first line we used the triangle inequality, and to go to the second line, we used the fact that $\Vert \mathcal{E}(\rho) - \mathcal{E}(\sigma) \Vert_1 \leq \Vert \rho - \sigma \Vert_1$ for any channel $\mathcal{E}$ (for a proof, see \cite[Exercise 9.1.9]{Wilde:2011npi}).
This completes the second step of the proof.

For the third step, we let $\Oh_A := \mathcal{R}^\dagger[\phi_a]$ and calculate:
\begin{align*}
\left| \langle \Oh_A \rangle_{J\rho J^\dagger} - \langle \phi_a \rangle_\rho  \right| &= \left| \, \Tr\left[ \mathcal{R}^\dagger[\phi_a] (J \rho J^\dagger)_A \right] - \Tr [\phi_a \rho_a] \, \right| \\
& = \left| \, \Tr\left[ \phi_a \mathcal{R}[(J \rho J^\dagger)_A] \right] - \Tr [\phi_a \rho_a]  \, \right| \\
&=  \left| \, \Tr\left[ \phi_a \left(\mathcal{R}[(J \rho J^\dagger)_A] -  \rho_a \right) \right] \, \right| \\
&\leq \left\Vert \, \phi_a \left(\mathcal{R}[(J \rho J^\dagger)_A] -  \rho_a \right) \, \right\Vert_1 \\
&\leq \left\Vert\mathcal{R}[(J \rho J^\dagger)_A] -  \rho_a \right\Vert_1 \, \Vert \phi_a \Vert \\
&\leq \delta \Vert \phi_a \Vert
\end{align*}
The only ``new'' ingredient that we used in these manipulations was Holder's inequality, $\Vert XY \Vert_1 \leq \Vert X \Vert_p \Vert Y \Vert_q$ for $p^{-1} + q^{-1} = 1$, to go from the fourth line to the fifth line (as well as the fact that $q \rightarrow \infty$ coincides with the operator norm).
We therefore obtain an accurate reconstruction of $\phi_a$ in terms of an operator $\Oh_A$ that only has support on $A$.

\section{Conclusion}
\label{sec:conc}

These notes introduced a handful of core ideas in quantum information through the lens of quantum channels.
In that sense, \Sec{sec:channels} was the core part of these notes, where we carefully defined what a channel is as well as certain channel properties.
In particular, we drew on the relative entropy machinery that we developed in \Sec{sec:QIbasics} to characterize a channel as a process that degrades distinguishability.
This characterization would prove key to understanding quantum error correction as a channel in \Sec{sec:QEC}, where we viewed encoding and noise as a channel that we attempt to reverse through a decoding channel.
We initially introduced universal recovery channels to this end, but they subsequently played a crucial role in \Sec{sec:holo} in interpreting bulk reconstruction in AdS/CFT as a quantum error correcting code.

The topics that we covered were chosen with an eye towards applications in high energy physics, particularly within the AdS/CFT correspondence, and so you should be well-equipped now to embark on further studies.
For example, the question of how one recovers information from a black hole can be thought of as an attempt to reverse a channel, and tools that we saw, like the Petz map, are showing up in some of the most recent studies of this problem \cite{Almheiri:2020cfm}.
It's an exciting time to be studying quantum information in quantum gravity.

\newpage
\begin{center} 
{\bf Acknowledgments}
\end{center}
\noindent 
I am grateful to the Modave Organizing Committee for organizing this school and for giving me the opportunity to attend as a lecturer, and to Kwinten Fransen for carefully proofreading these notes and handling the editorial aspects of publication.
I am also grateful to the other attendees and lecturers, whose contributions and interactions made the school a stimulating experience.
I would like to thank John Preskill for giving his permission to include Exercises 2 and 3 in these notes.
I am a Postdoctoral Fellow (Fundamental Research) of the Research Foundation -- Flanders (Fonds Wetenschappelijk Onderzoek), File Number 12ZL920N, and this work was supported by this fellowship.
\newpage

\section{Exercises}
\label{sec:exercises}

The purpose of these exercises is to give you a chance work with some of the tools that were introduced in these notes while filling in technical details.
Some of the exercises are based on homework problems that I had to solve when I was a student, and I am sure that that these problems or variations on them are still in use.
For this reason, I have not included solutions to the exercises.
Even so, if generations of students have made it through these problems in the past, I am sure that you will be able to do the same!

~

\noindent {\bf Exercise 1. The Schmidt decomposition}

\noindent Let $\Hil_{AB}$ be a separable Hilbert space, i.e. it admits a countable basis of orthonormal eigenvectors.
Furthermore, suppose that $\Hil_{AB}$ factorizes into the tensor product $\Hil_{AB} = \Hil_A \otimes \Hil_B$, and let $\ket{\psi}_{AB} \in \Hil_{AB}$.
We can always write
\begin{equation}
\ket{\psi}_{AB} = \sum_i \sum_\mu a_{i\mu} \ket{i}_A \ket{\mu}_B
\end{equation}
where $\{ \ket{i}_A \}$ and $\{ \ket{\mu}_B \}$ are orthonormal bases for $\Hil_A$ and $\Hil_B$, respectively.
For each $i$, let us define the vector $\ket{\tilde i}_B = \sum_{\mu} a_{i\mu} \ket{\mu}_B$, so that
\begin{equation} \label{eq:preSchmidt}
\ket{\psi}_{AB} = \sum_i \ket{i}_A \ket{\tilde i}_B.
\end{equation}
Note that the $\ket{\tilde i}_B$ need not be normalized nor orthogonal.

~

\noindent {\bf a)} Suppose that $\{ \ket{i}_A \}$ is the basis in which $\rho_A = \Tr_B \ketbra{\psi}{\psi}_{AB}$ is diagonal, and let the set $S$ label the non-zero eigenvalues of $\rho_A$, i.e. $p_i \neq 0 \Leftrightarrow i \in S$.
In other words,
\begin{equation} \label{eq:Astate}
\rho_A = \sum_{i \in S} p_i \ketbra{i}{i}_A .
\end{equation}
Starting from Eq.~\eqref{eq:preSchmidt}, compute $\rho_A$ by taking the partial trace over $B$ and show that
\begin{equation} \label{eq:preAstate}
\rho_A = \sum_i \sum_{i'} \braket{\tilde{i}'}{\tilde i}_B \ketbra{i}{i'}_A .
\end{equation}

\noindent {\bf b)} Compare Eqs.~\eqref{eq:Astate} and \eqref{eq:preAstate}.
What do you conclude about the overlap $\braket{\tilde{i}'}{\tilde i}$?
Use this to write down a set of orthonormal vectors in $B$.

~

\noindent {\bf c)} Write down $\ket{\psi}_{AB}$ using the basis $\{ \ket{i} \}_A$ and the orthonormal set of vectors in $B$ that you found above.
What are the eigenvalues of $\rho_B$?

~

\emph{Note:} This important result is known as the \emph{Schmidt decomposition}.
Any bipartite pure state $\ket{\psi}_{AB}$ can be written in the form
\begin{equation}
\ket{\psi}_{AB} = \sum_j \sqrt{p_j} \ket{\phi_j}_A \ket{\chi_j}_B,
\end{equation}
where the vectors $\ket{\phi_j}_A$ and $\ket{\chi_j}_B$ are orthonormal in $A$ and $B$, separately.
Note that this decomposition is state-dependent.
In general, if $\ket{\omega}_{AB}$ is some other state, then it will not have such a decomposition in terms of the same vectors.

~

\noindent {\bf Exercise 2. Distinguishability via the trace norm}

\noindent {\it Adapted with permission from \href{http://theory.caltech.edu/~preskill/ph219/chap2_13.pdf}{Exercise 2.7 of J. Preskill, \emph{Lecture Notes for Ph219/CS219: Quantum Information and Computation}, Chapter 2 (2013 edition)}.}

~

In many cases, we would like to be able to meaningfully quantify how ``close'' two quantum states are to each other.
For example, if we are trying to correct errors made during a quantum computation, we would like to quantify how close the post-recovery state is to the original state.
In this problem, we will see why the 1-norm is a good measure of closeness.

~

Consider two quantum states described by density operators $\rho$ and $\tilde \rho$ in a $N$-dimensional Hilbert space, and consider the complete orthogonal measurement $\{E_a : a = 1, 2, \dots, N\}$, where the $E_a$'s are one-dimensional projectors satisfying
\begin{equation} \label{eq:completeness}
\sum_{a=1}^N E_a = I. 
\end{equation}
When the measurement is performed, outcome $a$ occurs with probability $p_a = \Tr \rho E_a$ if the state is $\rho$ and with probability $\tilde p_a = \Tr \tilde{\rho} E_a$ if the state is $\tilde \rho$.

The (normalized) $L^1$ distance between the two probability distributions is defined as
\begin{equation}
d(p,\tilde{p}) \equiv \Vert p - \tilde{p} \Vert_1 \equiv \frac{1}{2} \sum_{a=1}^N |p_a - \tilde p_a|.
\end{equation}
This distance is zero if the two distributions are identical, and attains its maximum value of one if the two distributions have support on disjoint sets.

~

\noindent {\bf a)} Show that
\begin{equation} \label{eq:bound}
d(p,\tilde{p}) \leq \frac{1}{2} \sum_{i=1}^N |\lambda_i|,
\end{equation}
where the $\lambda_i$'s are the eigenvalues of the Hermitian operator $\rho - \tilde{\rho}$.
\textit{Hint:} Working in the basis in which $\rho - \tilde{\rho}$ is diagonal, find an expression for $|p_a - \tilde{p}_a|$, and then find an upper bound on $|p_a - \tilde{p}_a|$.
Finally, use the completeness property Eq.~\eqref{eq:completeness} to bound $d(p,\tilde{p})$.

~

\noindent {\bf b)} Find a choice for the orthogonal projectors $\{E_a\}$ that saturates the upper bound Eq.~\eqref{eq:bound}.

~

Define a distance $d(\rho,\tilde{\rho})$ between density operators as the maximal $L^1$ distance between the corresponding probability distributions that can be achieved by any orthogonal measurement.
From the results of (a) and (b), we have found that
\begin{equation}
d(\rho,\tilde \rho) = \frac{1}{2} \sum_{i=1}^N |\lambda_i|.
\end{equation}

~

\noindent {\bf c)} The trace norm, or Schatten 1-norm $\Vert A \Vert_1$ of an operator $A$ is defined as
\begin{equation}
\Vert A \Vert_1 \equiv \Tr \left[ (A^\dagger A)^{1/2} \right].
\end{equation}
How can the distance $d(\rho,\tilde \rho)$ be expressed as the $1$-norm of an operator?

~

Now suppose that the states $\rho$ and $\tilde \rho$ are pure states $\rho = \ketbra{\psi}{\psi}$ and $\tilde{\rho} = \ketbra{\tilde \psi}{\tilde \psi}$.
If we adopt a suitable basis in the space spanned by the two vectors, and appropriate phase conventions, then these vectors can be expressed as
\begin{equation}
\ket{\psi} = \left( \begin{array}{c} \cos \theta/2 \\ \sin \theta/2 \end{array} \right) \qquad \ket{\tilde \psi} = \left( \begin{array}{c} \sin \theta/2 \\ \cos \theta/2 \end{array} \right) .
\end{equation}

~

\noindent {\bf d)} Express the distance $d(\rho,\tilde \rho)$ in terms of the angle $\theta$.

~

\noindent {\bf e)} Express $\Vert \ket{\psi} - \ket{\tilde \psi} \Vert^2$ (where $\Vert \cdot \Vert$ denotes the Hilbert space norm, i.e., the 2-norm $\Vert \ket{\psi} \Vert = \sqrt{\braket{\psi}{\psi}}$) in terms of $\theta$, and by comparing with the results of (d), derive the bound
\begin{equation}
d(\ketbra{\psi}{\psi},\ketbra{\tilde \psi}{\tilde \psi}) \leq \Vert \ket{\psi} - \ket{\tilde \psi} \Vert .
\end{equation}

~

\noindent {\bf f)} Why is $\Vert \ket{\psi} - \ket{\tilde \psi} \Vert$ \emph{not} a good measure of the distinguishability of the pure quantum states $\rho$ and $\tilde \rho$?
\textit{Hint:} Remember that quantum states are \textit{rays}.

~

\noindent {\bf Exercise 3. Positivity of relative entropy}

\noindent {\it Adapted with permission from \href{http://theory.caltech.edu/~preskill/ph219/chap10_6A.pdf}{Exercise 10.1 of J. Preskill, \emph{Lecture Notes for Ph219/CS219: Quantum Information and Computation}, Chapter 10 (2018 edition)}.}

~

\noindent {\bf a)} Show that $\log x \leq x-1$ for all positive real numbers, with equality if and only if $x = 1$.

~

\noindent {\bf b)} The classical relative entropy of a probability distribution $\{p(x)\}$ relative to $\{q(x)\}$ is defined as
\begin{equation}
H(p \, \Vert \, q) = \sum_x p(x) \left( \log p(x) - \log q(x) \right),
\end{equation}
for distributions such that $p(x)= 0$ if $q(x) = 0$, and where the sum is over $x$ such that $q(x) \neq 0$.
Show that
\begin{equation}
H(p \, \Vert \, q) \geq 0,
\end{equation}
with equality if and only if the distributions are identical. (\emph{Hint:} apply the inequality from (a) to $\log(q(x)/p(x))$.)

~

\noindent {\bf c)} The quantum relative entropy of the density operator $\rho$ with respect to $\sigma$ is
\begin{equation}
D(\rho \, \Vert \, \sigma) = \Tr\left[ \rho \log \rho - \rho \log \sigma \right],
\end{equation}
and it is well-defined provided $\mrm{ker}~\sigma \subseteq \mrm{ker}~\rho$.
Let $\{ p_i\}$ denote the eigenvalues of $\rho$ and $\{q_a\}$ denote the eigenvalues of $\sigma$. Show that
\begin{equation}
D(\rho \, \Vert \sigma) = \sum_i p_i \left( \log p_i - \sum_a D_{ia} \log q_a \right),
\end{equation}
where $D_{ia}$ is a doubly stochastic matrix.
Express $D_{ia}$ in terms of the eigenstates of $\rho$ and $\sigma$. (A matrix is doubly stochastic if its entries are nonnegative real numbers, where each row and each column sums to one.)

~

\noindent {\bf d)} Show that if $D_{ia}$ is doubly stochastic, then (for each $i$)
\begin{equation}
\log\left(\sum_a D_{ia} q_a \right) \geq \sum_a D_{ia} \log q_a ,
\end{equation}
with equality only if $D_{ia} = 1$ for some $a$.

~

\noindent {\bf e)} Show that
\begin{equation}
D(\rho \, \Vert \, \sigma) \geq H(p \, \Vert r ),
\end{equation}
where $r_i = \sum_a D_{ia} q_a$.

~

\noindent {\bf f)} Show that $D(\rho \, \Vert \, \sigma) \geq 0$, with equality if and only if $\rho = \sigma$.

~

\noindent {\bf Exercise 4. Cyclicity of the partial trace}

\noindent Let $V \in \mathcal{L}(\Hil_A,\Hil_B)$, $W \in \mathcal{L}(\Hil_B, \Hil_A)$, and $\tau \in \mathcal{L}(\Hil_B)$.
Show that $\Tr_A[W \tau V] = \Tr_B[V W \tau]$.

~

\noindent {\bf Exercise 5. Syndrome measurement in the 9-qubit code}

\noindent Circuits are a useful way of depicting a sequence of unitary operations.
For example, the following circuit depicts $U \ket{\psi}$.
\begin{figure}[ht]
\centering
\includegraphics[scale=1]{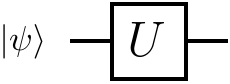}
\end{figure}

\noindent A horizontal line denotes a degree of freedom (such as a qubit), and boxes represent unitary operators.
Circuits are read left to right.
A box with the word ``measure'' denotes a measurement in the computational basis.

~

\noindent {\bf a)} Show that the following circuit measures $Z_1 Z_2$.

\begin{figure}[ht]
\centering
\includegraphics[scale=1]{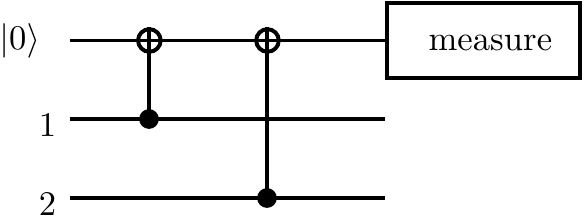}
\end{figure}

\clearpage

\noindent The two-qubit operator
\begin{figure}[ht]
\centering
\includegraphics[scale=1]{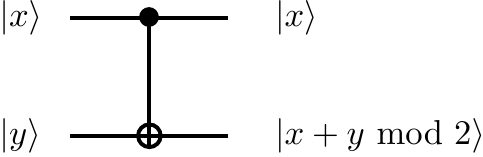}
\end{figure}

\noindent denotes the controlled-not, or CNOT operator.
Its action on two qubits is $\mrm{CNOT} \ket{x}\ket{y} = \ket{x} \ket{x + y ~ \mrm{mod}~2}$.

~

\noindent {\bf b)} Find a circuit that collectively and non-destructively measures $X_1 X_2 X_3 X_4 X_5 X_6$.
You may find that the single-qubit operator $H$ known as the \emph{Hadamard operator} is a helpful ingredient.
Its action is 
\begin{align*}
H \ket{0} &= \frac{1}{\sqrt{2}} \left( \ket{0} + \ket{1} \right) \equiv \ket{+} \\
H \ket{1} &= \frac{1}{\sqrt{2}} \left( \ket{0} - \ket{1} \right) \equiv \ket{-} .
\end{align*}

~

\noindent {\bf Exercise 6. Additivity of relative entropy}

\noindent Show that $D(\rho_A \otimes \chi_B \, \Vert \, \sigma_A \otimes \tau_B) = D(\rho_A \, \Vert \, \sigma_A) + D(\chi_B \, \Vert \, \tau_B)$.
You can assume that $\sigma_A$ and $\tau_B$ are full-rank (no zero eigenvalues) to avoid divergences in relative entropy.

~

\noindent \emph{Note:} If we think of $D$ as a measure of distinguishability, then the result above is clear.
The uncorrelated states in $B$ cannot influence the distinguishability of the states of $A$ and vice-versa.
This can also be viewed as a special case of monotonicity of relative entropy, $D(\rho_{AB} \, \Vert \, \sigma_{AB}) \geq D(\rho_A \, \Vert \, \sigma_A)$.

\clearpage

\bibliographystyle{utphys-modified}
\bibliography{modave_refs.bib}

\end{document}